\pdfoutput=1

\documentclass{article}

\PassOptionsToPackage{numbers,sort&compress}{natbib}

\usepackage[preprint]{neurips_2026}

\usepackage[utf8]{inputenc}
\usepackage[T1]{fontenc}
\usepackage{hyperref}
\usepackage{url}
\usepackage{booktabs}
\usepackage{amsfonts}
\usepackage{amsmath}
\usepackage{amssymb}
\usepackage{nicefrac}
\usepackage{microtype}
\usepackage{xcolor}
\usepackage{xcolor}
\usepackage{colortbl}
\usepackage{graphicx}
\usepackage{multirow}
\usepackage{array}
\usepackage{subcaption}
\usepackage{siunitx}
\usepackage{xspace}
\usepackage{enumitem}
\usepackage{algorithm}
\usepackage{algpseudocode}
\usepackage{cleveref}
\crefname{app}{Appendix}{Appendices}
\crefname{sec}{Section}{Sections}
\usepackage{cleveref}
\usepackage[figureposition=bottom,tableposition=top]{caption}

\graphicspath{{graphs/}{../figures/}{figures/}}

\makeatletter
\def\fps@figure{!t}
\def\fps@table{!t}
\makeatother

\newcommand{\best}[1]{\textbf{#1}}
\newcommand{\second}[1]{\underline{#1}}

\DeclareRobustCommand{\method}
{\texorpdfstring{\textsc{Sid-Mlp}}{Sid-Mlp}\xspace}
\DeclareRobustCommand{\methodpp}{\texorpdfstring{\textsc{Sid-Mlp++}}{Sid-Mlp++}\xspace}

\title{MLPs are Efficient Distilled \\ Generative Recommenders}

\author{
  Zitian Guo\textsuperscript{1},  
  Yupeng Hou\textsuperscript{1},  
  Clark Mingxuan Ju\textsuperscript{2}, 
  Neil Shah\textsuperscript{2}, 
  Julian McAuley\textsuperscript{1} 
  \\
  \textsuperscript{1} University of California, San Diego,    
  \textsuperscript{2} Snap Inc.\\
  \texttt{\{ztguo,yphou,jmcauley\}}@ucsd.edu,
  \\
  \texttt{\{mju,nshah\}}@snap.com
}

\begin{document}

\maketitle

\begin{abstract}
\label{sec:abstract}

Generative recommendation models employing Semantic IDs (SIDs) exhibit strong potential, yet their practical deployment is bottlenecked by the high inference latency of beam-expanded autoregressive decoding. In this work, we identify that standard attention-heavy Transformer decoders represent a structural overkill for this task: the hierarchical nature of SIDs makes prediction difficulty drops sharply after the first token, rendering repeated attention computations highly redundant.
Driven by this insight, we propose \method, a lightweight MLP-centric distillation framework that fundamentally simplifies the decoding paradigm for GR. Instead of executing complex, step-by-step attention mechanisms, our approach captures the global user context in a single operation, decoupled from sequential token prediction. We then distill the heavy autoregressive teacher into position-specific MLP heads, eliminating the dense attention overhead while preserving prefix and context dependencies. 
Extensive experiments demonstrate that \method matches the accuracy of teacher models while accelerating inference by $8.74\times$. Crucially, this distillation strategy can serve as a plug-and-play accelerator for different backbones and tokenizer settings.
Furthermore, we introduce \methodpp, extending our distillation framework to replace the Transformer encoder, unlocking further latency reductions.
Ultimately, our work reveals that decoder-side MLPs distillation is an effective acceleration path for structured SID recommendation, while full encoder replacement offers an additional speed--accuracy trade-off.
Our code is available at: ~\url{https://github.com/ztguo715/SID-MLP.git}.
\end{abstract}

\section{Introduction}

\label{sec:intro}

Generative recommendation (GR) approaches~\citep{dsi,tiger,lcrec,onerec} diverge from the conventional sequential recommendation paradigm~\citep{gru4rec,sasrec}, which models user histories with atomic item IDs; instead, each item is often represented by a semantic ID (SID), an ordered tuple of discrete tokens from a compact shared vocabulary. This formulation enables semantically similar items to share token prefixes (improving generalization) and avoid large embedding tables which scale with item cardinality (alleviating embedding sparsity). 
In this way, next-item prediction is framed as autoregressive SID generation, which has shown promising performance~\cite{tiger,lcrec,onerec,plum}.

However, the high inference latency of GR has hindered its deployment in production systems. Consider a GR model that represents each item with a 4-token SID. Unlike conventional models, which can produce top-$K$ predictions with a single model forward pass~\cite{gru4rec,sasrec}, GR models usually rely on beam search to generate multiple candidate SID sequences, with each sequence requiring four model forward passes due to the autoregressive generation mechanism. To achieve competitive performance, the beam size $B$ is often set even larger than $K$, leading to a total of roughly $4 \times B$ model forward passes during inference. This cost can be prohibitive in recommender systems, where latency is critical to user experience.

Developing broadly applicable methods for accelerating GR is non-trivial.
Speculative decoding~\citep{specdec,medusa,eagle} has been widely used to accelerate large language model inference. However, GR requires generating top-$K$ candidate sequences, which makes the verification step difficult to adapt, as it would need to verify and rank multiple candidate sequences simultaneously~\citep{atspeed}.
Multi-token, parallel, and non-autoregressive methods~\citep{rpg,nezha,nar4rec} remove the left-to-right decoding dependency,
but often require jointly fine-tuning the teacher model or relying on a specific tokenizer, limiting their applicability to general model architectures and other recent advances in the field.
Knowledge distillation (KD)~\citep{hinton,glnn} offers a promising pathway by transferring knowledge from a teacher model to a more efficient student model.
However, standard same-family KD (\emph{e.g.}, distilling into a smaller Transformer) only reduces model size while retaining much of the decoding time complexity.

Note that the decoding space of GR is typically well structured, leaving room for inference acceleration. Unlike large language models~\cite{zhao2023survey}, whose generations are open-ended and variable-length, GR models produce short and fixed-length outputs, such as 4-token SIDs in TIGER~\cite{tiger}, with a limited set of valid token sequences. As empirically observed in~\Cref{fig:perdigit}, once the first few tokens are determined, the number of valid continuations rapidly drops to only a few choices, far below the number of all theoretically possible combinations. This observation motivates a natural question: \emph{Is an attention-based Transformer decoder unnecessarily heavy for inference in GR?}

To this end, we investigate whether a simpler model architecture can be used for efficient GR inference. Our hypothesis is that a heavy model is still needed to learn from raw token sequences, but the structural regularity of SID generation enables a much cheaper approximation at inference time. We begin by replacing the original Transformer decoder with one of the simplest neural architectures: multilayer perceptrons (MLPs)~\cite{zhou2022fmlp,glnn}. However, naive MLPs do not perform well, since such regularities become useful only when the first few tokens are predicted correctly, as also demonstrated by our empirical results in~\Cref{tab:attn}.

Taking the above considerations together, we propose \textbf{\method}, an MLP-centric distillation framework for efficient GR inference. Given the importance of accurately predicting the first token, we retain a lightweight one-layer attention module to obtain a user-aware summary representation. A series of cascaded MLPs is then used to generate the remaining tokens sequentially. Since the decoding spaces at different positions are usually disjoint, we use a separate MLP with independent parameters for each position. To condition each position on the previously generated prefix, the input to each MLP is constructed by concatenating the embeddings of the generated prefix tokens.
Experiments on public benchmarks demonstrate that \method matches Transformer-based teacher model's performance with an average of $8.74\times$ speedup. We further introduce \textbf{\methodpp}, which extends the distillation to the encoder side and increases the speedup to $10.25\times$ while maintaining competitive performance.

\begin{table*}[!t]
\centering
\begin{minipage}[t]{0.38\textwidth}
  \vspace{0pt}
  \centering
  \includegraphics[width=\linewidth]{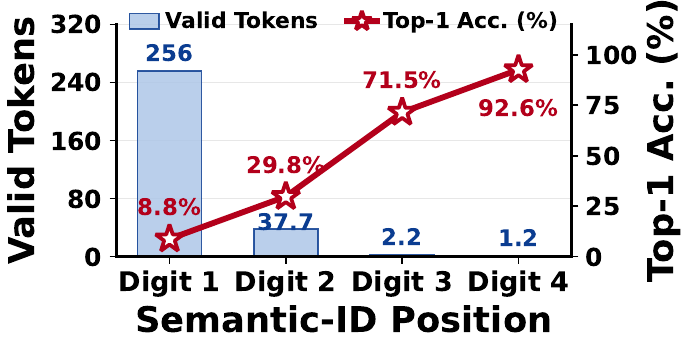}
  \captionof{figure}{SID search-space collapse on Instruments. Average Valid next-token
  choices and top-1 accuracy.}
  \label{fig:perdigit}
\end{minipage}\hfill
\begin{minipage}[t]{0.29\textwidth}
  \vspace{0pt}
  \centering
  {\setlength{\tabcolsep}{2.5pt}\renewcommand{\arraystretch}{0.90}\scriptsize
  \captionof{table}{Decoder depth redundancy on Instruments. Teacher generates first $m$ digits; a 1-layer decoder predicts the remaining $4{-}m$.}
  \label{tab:nmode}
  \begin{tabular}{lcc}
    \toprule
    Variant & N@10 & $\Delta$ vs teacher \\
    \midrule
    TIGER  & 0.0323 & --- \\
    $m{=}3$ & 0.0323 & $\pm 0.0\%$ \\
    $m{=}2$ & 0.0320 & $-0.9\%$ \\
    $m{=}1$ & 0.0320 & $-0.9\%$ \\
    $m{=}0$ & 0.0318 & $-1.5\%$ \\
    \bottomrule
  \end{tabular}\par}
\end{minipage}\hfill
\begin{minipage}[t]{0.29\textwidth}
  \vspace{0pt}
  \centering
  {\setlength{\tabcolsep}{2.5pt}\renewcommand{\arraystretch}{0.90}\scriptsize
  \captionof{table}{Attention ablation on the $m{=}0$ 1-layer decoder on Instruments. SA and CA denote self-attention and cross-attention, respectively.}
  \label{tab:attn}
  \begin{tabular}{cccc}
    \toprule
    SA & CA & N@10 & $\Delta$ vs baseline \\
    \midrule
    \(\checkmark\) & \(\checkmark\) & 0.0318 & --- \\
    \(\times\) & \(\checkmark\) & 0.0310 & $-5.4\%$ \\
    \(\checkmark\) & \(\times\) & 0.0131 & $-60.0\%$ \\
    \(\times\) & \(\times\) & 0.0111 & $-65.1\%$ \\
    \bottomrule
  \end{tabular}\par}
\end{minipage}
\end{table*}
\vspace{-0.4\baselineskip}
\section{Motivation Study}
\label{sec:motivation}

\subsection{Preliminary: Generative Recommendation}
\label{sec:prelim-tiger}

\textbf{Task formulation.} Let $\mathcal{U}$ and $\mathcal{V}$ denote the sets of users and items, respectively. For a user $u \in \mathcal{U}$, their historical interaction sequence is represented as $\mathbf{X}_u = [v_1, v_2, \dots, v_n]$, where $v_i \in \mathcal{V}$. Each item $v$ is uniquely mapped to a Semantic ID, which is a tuple of $L$  
discrete tokens: $c^{(v)} = (c_1, c_2, \dots , c_L)$. For ease of notation, we assume $L=4$ in our subsequent examples, yielding $c^{(v)} = (c_1, c_2, c_3, c_4)$. In TIGER, the first three tokens are residual quantization codes of item embeddings from a shared codebook $\mathcal{C}$ ($|\mathcal{C}| = 256$), and the fourth is a conflict avoidance token. TIGER serializes $\mathbf{X}_u$ into an encoder token sequence $\mathbf{s}_u$ containing $S_u$ non-padding tokens. A multi-layer T5 encoder-decoder then generates the next item's SID by factorizing the joint probability autoregressively: $p_\theta(c^{(v)} \mid \mathbf{s}_u) = \prod_{j=1}^{L} p_\theta(c_j \mid c_{<j}, \mathbf{s}_u)$.

\textbf{The autoregressive mechanism and latency explosion.} Autoregressive SID generation relies on beam search. For an SID of length $L$, beam size $B$, and an $N$-layer decoder, a cached decoder still runs about $L B N$ decoder-block evaluations. Across all digit steps, this includes $\mathcal{O}(B N L^2)$ prefix self-attention work and $\mathcal{O}(L B N S_u)$ cross-attention reads over the encoder states. Since digit $c_j$ depends on prefix $c_{<j}$, SID positions cannot be decoded in parallel, making this sequential process the main inference bottleneck.

\subsection{Motivation: Rethinking the Decoder's Necessity}
\label{sec:motivation-study}

Given the inference latency caused by autoregressive decoding, we investigate the structural necessity of the full Transformer decoder through data-level and architecture-level analyses.

\textbf{SID branching concentrates uncertainty in early digits.}\label{sec:motivation-perdigit} We first characterize the inherent difficulty of the SID prediction task. As shown in Figure~\ref{fig:perdigit}, the valid codebook branching factor given a valid prefix drops sharply. On the Instruments dataset, the average branching factor drops from $256\!\to\!{\sim}38\!\to\!2.2\!\to\!{\sim}1.2$. Per-dataset statistics in Appendix~\ref{app:codebook} show the same collapse pattern on Scientific and Games. 
The recommendation problem remains history-dependent, but decoder-side uncertainty is highly uneven: most branching is concentrated in the early digits, while later valid digits are strongly constrained by prefixes. For $c_3$ and $c_4$, the full $N$-layer decoder spends computationally expensive forward passes on a nearly collapsed search space.

\textbf{Decoder depth is heavily over-parameterized.}\label{sec:motivation-nmode} Given this sharp drop in prediction difficulty, we investigate how much decoder depth is actually utilized. We train a minimal 1-layer student decoder distilled from the frozen TIGER teacher (training recipe in~\Cref{sec:exp-setup}). The teacher generates the first $m \in \{0,1,2,3\}$ digits, and the student predicts the remaining $4-m$ digits. As shown in Table~\ref{tab:nmode}, a 1-layer student perfectly matches the 4-layer teacher when predicting only $c_4$ ($m{=}3$), and loses a mere $1.5\%$ NDCG@10 even when predicting all four digits ($m{=}0$). This suggests that much of the decoder depth is not essential for SID-level ranking once the teacher's encoder representation and distillation signal are available. 
We focus on distillation because the accuracy gap between scratch-training and distillation widens as the architecture simplifies into MLPs (Appendix~\ref{app:scratch}).

\textbf{User-history context must be preserved.}
Since decoder depth can be safely reduced, we investigate whether we can further decompose the remaining Transformer block to remove redundant internal components. Table~\ref{tab:attn} evaluates the impact of removing specific attention modules from the $m{=}0$ 1-layer decoder. Removing prefix self-attention causes a $5.4\%$ NDCG@10 drop, indicating the short SID prefix still needs a position-aware representation. In contrast, removing cross-attention severs the decoder's access to the encoded user history, causing a massive $60.0\%$ drop. Removing both yields a $65.1\%$ loss, proving that a purely context-free MLP decoder is insufficient. Cross-dataset replications confirm this pattern (Appendix~\ref{app:nmode}).

\textbf{Target: Bridging efficiency and context-awareness.} 
These ablations present a clear structural dilemma. A context-free MLP decoder offers high inference efficiency by eliminating attention overhead, but fails severely because it loses both the fine-grained historical alignment and the position-aware prefix dependency. Conversely, the standard Transformer decoder captures this necessary context but remains impractically slow: even with caching, each active beam still runs decoder-block updates and cross-attention reads at every digit. Any successful decoder replacement must explicitly preserve both the dynamic historical context and the prefix structure while avoiding these repeated computations. \emph{Can we achieve MLP-level inference speeds while retaining the structural benefits of both self- and cross-attention?} This objective motivates our design of \method in~\Cref{sec:method}.

\section{\method: Distilling Autoregressive Transformer Decoding to MLPs}
\label{sec:method}

\subsection{Overview of \method}
\label{sec:method-overview}

\begin{figure}[!t]
  \centering
  \includegraphics[width=\linewidth]{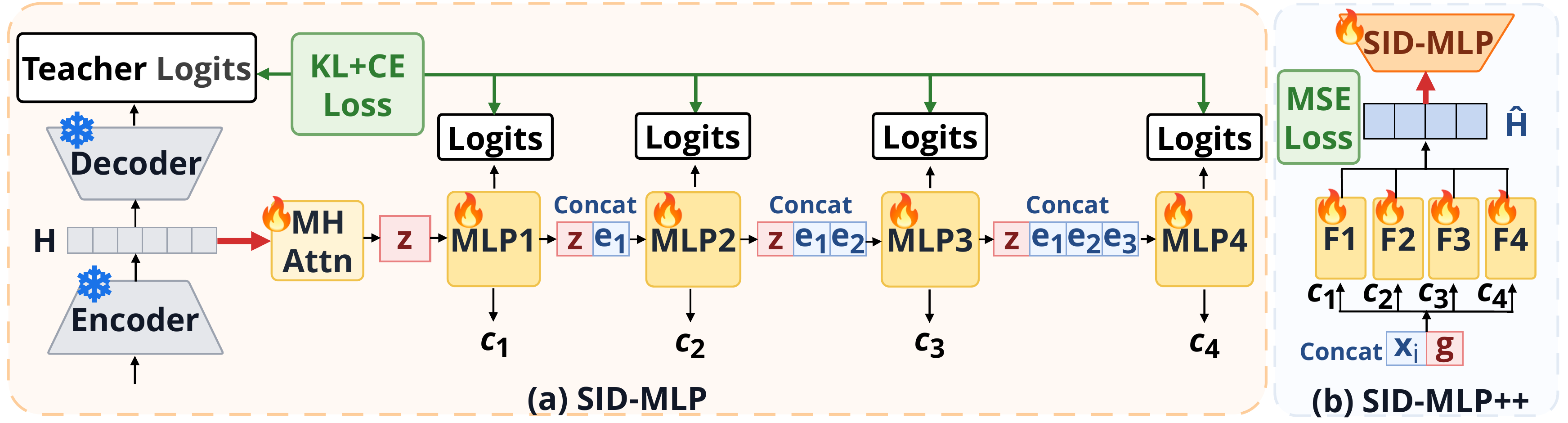}
  \caption{\textbf{Architecture.} 
  The architecture is composed of two components: \method (shaded yellow, left) and \methodpp (shaded blue, right). 
  MH Attn is multi-head attention, $\mathbf{e}_i = e(c_i)$ represents the embedding of prefix $c_i$. Snowflakes mark frozen modules, and flames mark trainable modules.}
  \label{fig:arch}
\end{figure}

To resolve the autoregressive latency and heavy attention overhead identified in~\Cref{sec:motivation}, we propose \textbf{\method}. 
Our core idea is to replace the redundant attention operations in autoregressive generative recommendation with lightweight MLP heads. As illustrated in \Cref{fig:arch}, we train these heads via knowledge distillation from a frozen Transformer teacher. This enables \method to achieve highly efficient inference while preserving the teacher's strong performance.

In this section, we first detail the extraction of a global user context from the frozen encoder (\Cref{sec:method-attn}), then introduce the prefix-conditioned MLP heads (\Cref{sec:method-heads}). Finally, we describe how we transfer the teacher's knowledge to MLPs via knowledge distillation (\Cref{sec:method-loss}). We conclude with an extension, \methodpp, that distills the encoder for further acceleration (\Cref{sec:method-enc}).

\vspace{-4pt}

\subsection{One-Shot Multi-head Attention Context}
\label{sec:method-attn}

The motivation study shows that encoder-conditioned user context is essential. \method therefore keeps an explicit multi-head attention readout, but computes it once outside the beam loop.

Let $\mathbf{H}_u=E_T(\mathbf{s}_u)\in\mathbb{R}^{S_u\times d_h}$ be the frozen TIGER encoder states for user $u$. 
We mean-pool these states, project the result into a query, and use a multi-head attention block to produce a context vector that is reused across all SID digit steps:
\begin{equation}
    \mathbf{q} = \mathrm{MeanPool}(\mathbf{H}_u)\,W_q,\quad
    \tilde{\mathbf{z}} = \mathrm{LN}\big(\mathbf{q} + \mathrm{MHA}(\mathbf{q}, \mathbf{H}_u, \mathbf{H}_u)\big),\quad
    \mathbf{z} = \mathrm{LN}\big(\tilde{\mathbf{z}} + \mathrm{FFN}(\tilde{\mathbf{z}})\big).
  \label{eq:global-context}
\end{equation}
where MHA is multi-head attention, LN is layer normalization, and FFN is feed-forward network.
   
\textbf{Complexity comparison.} Standard decoders (\emph{e.g.}, TIGER) compute cross-attention at every digit step, beam, and decoder layer, yielding an $\mathcal{O}(L B N S_u)$ historical alignment cost. Conversely, \method extracts the global context exactly once using the aggregated query $\mathbf{q}$. By caching $\mathbf{z}$, we entirely remove cross-attention from the beam search loop, collapsing this alignment cost to $\mathcal{O}(S_u)$.

\subsection{Prefix Concatenation and Per-Digit MLP Heads}
\label{sec:method-heads}
After $\mathbf{z}$ is cached, each digit prediction only requires the current SID prefix. \method completely removes standard self-attention. Instead, at step $t$, each beam simply concatenates the context $\mathbf{z}$ with $t-1$ frozen token embeddings $e(\cdot)$ retrieved from the teacher.  
Because SID digits have different output distributions, each position $t \in \{1, \dots, L\}$ uses a dedicated 1-hidden-layer MLP head $f_t$ to predict the next token: 
\begin{equation}
    \mathbf{p}_t = \big[\,\mathbf{z}\,;\, e(c_1)\,; \dots;\, e(c_{t-1})\,\big] \in \mathbb{R}^{d_h+(t-1)d_e},\quad
    \boldsymbol{\ell}_t = f_t(\mathbf{p}_t) \in \mathbb{R}^{C}.
    \label{eq:digit-head}
\end{equation}
where $[\cdot;\cdot]$ is concatenation, $c_1,\ldots,c_{t-1}$ are the available prefix digits, $d_e$ is the frozen embedding dimension, and $C$ (typically 256) is the codebook size. The input dimension of $f_t$ grows with $t$ and reaches at most $d_h+(L-1)d_e$.  

\textbf{Complexity comparison.} Standard decoders process prefixes via step-wise self-attention, yielding an $\mathcal{O}(B N L^2)$ cost alongside incremental KV-cache updates. Conversely, \method scores all active beams via batched dense MLP operations over static concatenated embeddings. Prefix processing thus becomes an update-free, constant-size operation per beam.

\subsection{Distillation Training and Inference}
\label{sec:method-loss}

We train \method via offline knowledge distillation from the frozen teacher. Only the single multi-head attention block and the MLP heads are trainable; the TIGER encoder is frozen. Each MLP head predicts the codebook slice for its digit, and the teacher logits are sliced to the same support. 
During teacher-forced training, \Cref{eq:digit-head} takes the ground-truth SID prefix. Let $\tilde{\boldsymbol{\ell}}_t$ denote the teacher logits and $c_t^\star$ the ground-truth codebook index. We train each digit with a Kullback--Leibler distillation term plus cross-entropy:
\begin{equation}
  \mathcal{L}_t = \alpha \tau^2 D_{\mathrm{KL}}\Big(\sigma(\tilde{\boldsymbol{\ell}}_t/\tau)\,\|\,\sigma(\boldsymbol{\ell}_t/\tau)\Big)
  + (1-\alpha)\mathrm{CE}(\boldsymbol{\ell}_t, c_t^\star)
  \label{eq:loss}
\end{equation}
where $\tau$ is the distillation temperature, $\sigma$ is softmax, and $\alpha \in [0,1]$ balances teacher mimicry and task grounding. The total loss is $\mathcal{L} = \sum_{t=1}^{L}\mathcal{L}_t$.

During inference, the context vector $\mathbf{z}$ is computed exactly once. Beam search then evaluates the MLP heads sequentially, batching all active prefixes at each digit step. We use constrained beam search: a valid-prefix mask built from the fixed item-to-SID mapping during tokenization stage, removes invalid prefixes before expansion. Algorithm~\ref{alg:sidmlp} summarizes the complete pipeline.

\begin{algorithm}[t]
\caption{\textbf{\method}: Per-digit MLP decoder distillation and beam-search inference.}
\label{alg:sidmlp}
\begin{algorithmic}[1]
\Require Frozen encoder $E_T$ and token embeddings $e(\cdot)$; serialized encoder tokens $\mathbf{s}_u$; 256-way teacher logits $\{\tilde{\boldsymbol{\ell}}_t\}_{t=1}^{L}$; training records; beam size $B$.
\Ensure Trained \method; top-$K$ item list during inference.
\Statex \textbf{Phase 1: Training Process (Teacher-Forcing)}
\For{each minibatch $(\mathbf{s}_u, \mathbf{c}^\star, \tilde{\boldsymbol{\ell}}_{1:L})$}
  \State $\mathbf{H}_u \gets E_T(\mathbf{s}_u)$
  \State $\mathbf{z} \gets \mathrm{GlobalContext}(\mathbf{H}_u)$ \Comment{Compute context \emph{once} per user}
  \For{$t = 1, \ldots, L$} \textbf{in parallel} \Comment{Valid due to known ground-truth prefix}
    \State $\mathbf{p}_t \gets [\mathbf{z}; e(c_1); \dots; e(c_{t-1})]$
    \State $\boldsymbol{\ell}_t \gets f_t(\mathbf{p}_t)$ \Comment{Output 256-way logits over codebook slice}
  \EndFor
  \State Compute $\mathcal{L}$ using \Cref{eq:loss} and update the trainable \method modules via AdamW
\EndFor
\Statex \textbf{Phase 2: Beam-Search Inference (Sequential with Batched Beams)}
\State $\mathbf{H}_u \gets E_T(\mathbf{s}_u)$
\State $\mathbf{z} \gets \mathrm{GlobalContext}(\mathbf{H}_u)$ \Comment{Executed \emph{only once} per user}
\State Initialize active beams $\mathcal{B}_0 \gets \{ \text{empty prefix} \}$
\For{$t = 1, \ldots, L$} \Comment{Strictly sequential across digit steps}
  \State \textbf{Batch compute} $\boldsymbol{\ell}_t \gets f_t([\mathbf{z}; e(c_1); \dots; e(c_{t-1})])$ for all $B$ prefixes in $\mathcal{B}_{t-1}$
  \State Apply the valid-prefix mask, expand beams within the codebook slice, and retain top-$B$ candidates into $\mathcal{B}_t$
\EndFor
\State \Return top-$K$ items decoded from the final beam $\mathcal{B}_L$
\end{algorithmic}
\end{algorithm}

\subsection{Extension: Encoder Distillation (\methodpp)}
\label{sec:method-enc}

\method removes the Transformer decoder stack but retains the teacher encoder. \methodpp extends distillation to this encoder. It replaces the teacher encoder with an MLP encoder and produces encoder hidden states $\hat{\mathbf{H}}_u = G_\phi(\mathbf{s}_u)\in\mathbb{R}^{S_u\times d_h}$; the following \method structure is unchanged.

\textbf{\methodpp architecture.} The encoder input is the serialized user-history sequence $[\mathtt{user}, c_1^{(v_1)}, \dots, c_L^{(v_1)}, \dots, \mathtt{EOS}]$. We assign four role-specific MLPs ($F_1, F_2, F_3, F_4$) to process the tokens, where $F_4$ handles $c_4$, \texttt{user}, and \texttt{EOS}, and the others handle their respective $c_j$ digits. Let $a_i \in \{1,2,3,4\}$ denote the role index for token $i$. The initial state $\mathbf{x}_{u,i}^{(0)}$ combines the frozen token embedding and learnable position embeddings. To replace self-attention, we achieve global context modeling at each layer $r$ by mean-pooling all non-padding token states into a global vector $\mathbf{g}_u^{(r)}$. Each token state $\mathbf{x}_{u,i}^{(r)}$ is then concatenated with this global vector and passed through its role-specific MLP to form a residual update:
\begin{equation}
  \mathbf{g}_u^{(r)} =
  \frac{1}{S_u}\sum_{i=1}^{S_u}\mathbf{x}_{u,i}^{(r)},\quad
  \mathbf{x}_{u,i}^{(r+1)} = \mathbf{x}_{u,i}^{(r)} + F_{a_i}\big([\mathbf{x}_{u,i}^{(r)};\mathbf{g}_u^{(r)}]\big).
  \label{eq:methodpp-encoder}
\end{equation}

\textbf{Two-stage encoder distillation.} Because direct logit distillation exhibits optimization instability for the encoder (ablation details in Appendix~\ref{app:encoder_ablation}), we adopt a two-stage process:
\begin{enumerate}[leftmargin=*, topsep=2pt, itemsep=2pt]
  \item \textbf{Stage 1 (Representation matching):} We pre-train the \methodpp encoder to match the frozen teacher's encoder states $\mathbf{H}_u$ using Mean Squared Error (MSE) loss.
  \item \textbf{Stage 2 (Logit distillation):} We freeze the student encoder, pass its output $\hat{\mathbf{H}}_u$ to the \method decoder, and train the decoder using the per-digit KL+CE loss (\Cref{eq:loss}).
\end{enumerate}

At inference, \methodpp replaces the teacher encoder states $\mathbf{H}_u=E_T(\mathbf{s}_u)$ with student states $\hat{\mathbf{H}}_u=G_\phi(\mathbf{s}_u)$. The cached context computation and sequential MLP heads are unchanged.
\section{Experiments}
\label{sec:experiments}

\subsection{Setup}
\label{sec:exp-setup}

\textbf{Datasets and Baselines.} We instantiate \method on \textbf{TIGER}~\citep{tiger}, a T5-based autoregressive generative recommender where items are tokenized into 4-digit semantic IDs. \method freezes the TIGER encoder and replaces decoder-side generation with prefix-conditioned MLP heads. We evaluate on three categories from the latest Amazon Reviews 2023 dataset~\citep{hou2024bridging}: Musical Instruments \textbf{(Instruments)}, Industrial \& Scientific \textbf{(Scientific)}, and Video Games \textbf{(Games)}. Full dataset statistics are in Appendix~\ref{app:datasets}. We organize baselines into three distinct groups.

\begin{enumerate}[label=\roman*, leftmargin=*, topsep=2pt, itemsep=2pt]
\item \emph{Teacher:} The TIGER model serves as our autoregressive upper bound, using a four-layer T5 to generate SIDs token-by-token via beam search. We report both the standard no-cache teacher TIGER and TIGER-kv, which enables the decoder key--value cache during generation. We further evaluate our method on the LC-Rec teacher, with experimental details in Appendix~\ref{app:lcrec}.

\item \emph{LLM-style Accelerators Ported to TIGER:} These baselines add draft/verification modules or jointly fine-tune TIGER, whereas \method keeps TIGER frozen. (1) \textbf{AtSpeed}~\citep{atspeed}, originally instantiated on LC-Rec~\citep{lcrec}, uses a compact draft model for speculative decoding and teacher verification; we port it with a compact T5 draft and evaluate strict verification (-S) and relaxed sampling (-R). (2) \textbf{EARN}~\citep{earn} uses boundary register tokens to retain early-layer history information and prunes other states; we port it to the TIGER encoder with joint fine-tuning. (3) \textbf{NEZHA}~\citep{nezha} uses SID placeholders and an autoregressive draft head with recurrent state updates for self-drafting; we feed the placeholders to the TIGER encoder and use the draft head instead of the T5 decoder, with joint fine-tuning. \Cref{tab:intro-struct} compares these models.

\item \emph{State space decoders:} These baselines test whether existing linear-recurrent sequence models can serve as lightweight replacements for the TIGER decoder. They keep the TIGER encoder frozen, feeding the encoder hidden states, a bridge token, and prefix embeddings into a causal SSM stack. The SSM outputs digit logits at these added positions and is trained with the same KL+CE objective as \method. We evaluate on (4) \textbf{GatedDeltaNet (GDN)}~\citep{gdn} and (5) \textbf{Mamba2}~\citep{mamba2}.
\end{enumerate}

\textbf{Evaluation \& Implementation.} We report Recall@$K$ and NDCG@$K$ ($K\in\{5,10\}$) over valid 4-digit SIDs. For the main \method row, performance metrics are averaged over random seeds \(\{42,43,44\}\); \Cref{app:seeds} reports mean\(\pm\)std and a competitive performance against the fixed TIGER-kv teacher. All methods use matched TIGER checkpoints. Throughput is end-to-end samples/s on the test split with batch size 32 and beam size \(B{=}50\); speedup is relative to TIGER-kv. Hyperparameters are in Appendix~\ref{app:hyperparams}; baseline adaptation and diagnostic details are in Appendices~\ref{app:atspeed-analysis}, \ref{app:baseline-adaptations}, and~\ref{app:nezha-repro}; hardware profiling details are in Appendix~\ref{app:hardware-profile}.

\subsection{Main Results}
\label{sec:exp-main}

\begin{table*}[t]
  \caption{\textbf{Main results.} Ranking metrics are mean values; \Cref{app:seeds} reports \method seed stability over \(\{42,43,44\}\). Tput is the averaged throughput across three datasets, and Spd. is the speedup relative to TIGER-kv. \best{Bold} = best, \second{underline} = second-best.}
  \label{tab:main}
  \centering
  \small
  \setlength{\tabcolsep}{3.2pt}
  \resizebox{\textwidth}{!}{%
  \begin{tabular}{p{1.98cm} rr cccc cccc cccc}
    \toprule
    & & & \multicolumn{4}{c}{\textbf{Instruments}} & \multicolumn{4}{c}{\textbf{Scientific}} & \multicolumn{4}{c}{\textbf{Games}} \\
    \cmidrule(lr){4-7}\cmidrule(lr){8-11}\cmidrule(lr){12-15}
    Method & Tput & Spd. & R@5 & R@10 & N@5 & N@10 & R@5 & R@10 & N@5 & N@10 & R@5 & R@10 & N@5 & N@10 \\
    \midrule
    TIGER  & 320 & 0.75$\times$ & 0.0386 & 0.0606 & 0.0252 & 0.0323 & 0.0295 & 0.0457 & 0.0191 & 0.0243 & \best{0.0612} & \second{0.0951} & \best{0.0403} & \best{0.0512} \\
    TIGER-kv  & 424 & 1.00$\times$ & 0.0386 & 0.0606 & 0.0252 & 0.0323 & 0.0295 & 0.0457 & 0.0191 & 0.0243 & \best{0.0612} & \second{0.0951} & \best{0.0403} & \best{0.0512} \\
    \midrule
    GDN  & 510 & 1.20$\times$ & 0.0385 & 0.0598 & 0.0253 & 0.0321 & 0.0278 & 0.0439 & 0.0183 & 0.0235 & 0.0585 & 0.0914 & 0.0380 & 0.0486 \\
    Mamba2  & 517 & 1.22$\times$ & 0.0388 & 0.0605 & 0.0256 & 0.0326 & \best{0.0300} & \second{0.0468} & \best{0.0193} & \second{0.0247} & 0.0589 & 0.0931 & 0.0386 & \second{0.0496} \\
    AtSpeed-R  & 94 & 0.22$\times$ & 0.0246 & 0.0375 & 0.0163 & 0.0204 & 0.0216 & 0.0320 & 0.0144 & 0.0178 & 0.0436 & 0.0637 & 0.0296 & 0.0361 \\
    AtSpeed-S  & 286 & 0.68$\times$ & 0.0386 & 0.0606 & 0.0252 & 0.0323 & 0.0295 & 0.0457 & 0.0191 & 0.0243 & 0.0612 & 0.0951 & 0.0403 & 0.0512 \\
    EARN  & 822 & 1.94$\times$ & 0.0383 & 0.0596 & 0.0250 & 0.0319 & 0.0288 & 0.0454 & 0.0186 & 0.0239 & 0.0584 & 0.0921 & 0.0377 & 0.0486 \\
    NEZHA  & 3{,}082 & 7.27$\times$ & 0.0371 & 0.0567 & 0.0245 & 0.0308 & 0.0254 & 0.0402 & 0.0164 & 0.0212 & 0.0560 & 0.0865 & 0.0368 & 0.0467 \\
    \midrule
    \rowcolor[gray]{0.93}\textbf{\method} & \second{3{,}706} & \second{8.74$\times$} & \best{0.0396} & \best{0.0620} & \best{0.0259} & \best{0.0332} & \second{0.0297} & \best{0.0472} & \best{0.0193} & \best{0.0250} & \second{0.0610} & \best{0.0953} & \second{0.0402} & \best{0.0512} \\
    \rowcolor[gray]{0.93}\textbf{\methodpp} & \best{4{,}347} & \best{10.25$\times$} & \second{0.0395} & \second{0.0612} & \second{0.0257} & \second{0.0327} & 0.0295 & 0.0459 & \second{0.0192} & 0.0244 & 0.0578 & 0.0916 & 0.0378 & 0.0486 \\
    \bottomrule
  \end{tabular}}
\end{table*}

\providecommand{\cmarkg}{\textcolor{green!55!black}{\checkmark}}
\providecommand{\xmarkr}{\textcolor{red!70!black}{\(\times\)}}
\begin{table}[!t]
  \caption{\textbf{Comparison of generative recommendation acceleration paradigms.} \emph{Plug-and-Play}: acts as a drop-in accelerator for already deployed models without requiring modifications to the original tokenizers, or fine-tuning of the base model. \emph{Verify-Free}: requires no target-model verification passes. \emph{Attn-Free Decoding}: isolates attention computation from the autoregressive loop, eliminating repeated self-attention and cross-attention during generation. \emph{Update-Free Beam}: executes beam expansion as batched matrix multiplications without maintaining KV-cache updates or recurrent hidden states.}
  \label{tab:intro-struct}
  \centering
  \footnotesize
  \setlength{\tabcolsep}{5.0pt}
  \renewcommand{\arraystretch}{1.05}
  \begin{tabular}{lcccc}
    \toprule
    Paradigm & Plug-and-Play & Verify-Free & Attn-Free Decoding & Update-Free Beam \\
    \midrule
    Speculative (AtSpeed~\citep{atspeed}) & \cmarkg & \xmarkr & \xmarkr & \xmarkr \\
    State Pruning (EARN~\citep{earn}) & \xmarkr & \cmarkg & \xmarkr & \xmarkr \\
    Self-Drafting (NEZHA~\citep{nezha}) & \xmarkr & \cmarkg & \cmarkg & \xmarkr \\
    \rowcolor[gray]{0.93}\textbf{\method (Ours)} & \cmarkg & \cmarkg & \cmarkg & \cmarkg \\
    \bottomrule
  \end{tabular}
  \vspace{-7pt}
\end{table}

\Cref{tab:main} compares \method against all baselines on the three categories. We highlight four findings:

\textbf{(1) \method is lossless without target-model verification.}
\method matches or exceeds TIGER NDCG@10 on all datasets while keeping the TIGER encoder frozen and requiring no target-model verification. This supports the motivation study: for short hierarchical SIDs, the teacher decoder's ranking behavior can be distilled into per-digit MLP heads as long as the student preserves the encoded user context and the ordered SID prefix. Additional hyperparameter analyses are in Appendix~\ref{app:hparam-figs}.

\textbf{(2) The speedup comes from removing repeated decoder work.}
\method reaches 3{,}706 samples/s (an 8.74$\times$ speedup over TIGER-kv), with a 95.7\% peak-memory reduction shown in Appendix~\ref{app:hardware-profile}. After one encoder pass and a multi-head attention readout, beam expansion relies solely on batched MLP projections. It eliminates decoder blocks, KV-cache updates, repeated attention reads, and draft verification. Distilling the encoder (\methodpp) further pushes the speedup to 10.25$\times$ with a minimal accuracy tradeoff.

\textbf{(3) Direct ports expose different bottlenecks.}
AtSpeed-S and AtSpeed-R are both slower than TIGER-kv ($0.68\times$ and $0.22\times$) because even our smallest single layer Transformer draft is too large relative to the 4.59M TIGER teacher, unlike the 68M-vs-7B LC-Rec setting~\citep{atspeed}. EARN reaches only $1.94\times$ because it only compresses the encoder states, leaving the autoregressive decoder blocks and per-step computation intact. Its quality also drops because TIGER's bidirectional encoder lacks the head/tail attention-sink pattern that motivates EARN's registers. NEZHA reaches $7.27\times$ by replacing the decoder with a draft-head path, but it still loses up to $15\%$ relative NDCG@10. \method is $1.20\times$ faster than NEZHA and preserves teacher quality. This highlights the architectural mismatch of directly porting LLM accelerators to GR.

\textbf{(4) SSMs are limited by recurrent memory updates overhead.}
Mamba2 (1.22$\times$) and GDN (1.20$\times$) are only slightly faster than TIGER-kv and much slower than \method despite their linear sequence complexity. In four-digit beam search, each step must update, fork, and gather recurrent states across active beams, so state movement dominates. \method is update-free after the context readout and scores all active prefixes with batched dense projections.

\subsection{Ablation Study}
\label{sec:exp-ablation}

\vspace{-4pt}

We ablate \method along three design axes. \Cref{tab:ablation} reports NDCG@10 and Recall@10 across Instruments, Scientific and Games; $\Delta$ columns show relative NDCG@10 change vs \method.

\begin{table}[!t]
  \caption{\textbf{Ablation study.} NDCG@10 and Recall@10 on three datasets; $\Delta\mathrm{NDCG}@10$ is relative to \method.}
  \label{tab:ablation}
  \centering
  \vspace{0.8mm}
  \tiny
  \setlength{\tabcolsep}{1.4pt}
  \renewcommand{\arraystretch}{0.76}
  \resizebox{0.90\linewidth}{!}{%
  \begin{tabular}{l ccc ccc ccc}
    \toprule
    & \multicolumn{3}{c}{\textbf{Instruments}} & \multicolumn{3}{c}{\textbf{Scientific}} & \multicolumn{3}{c}{\textbf{Games}} \\
    \cmidrule(lr){2-4}\cmidrule(lr){5-7}\cmidrule(lr){8-10}
    Variant & R@10 & N@10 & $\Delta$ & R@10 & N@10 & $\Delta$ & R@10 & N@10 & $\Delta$ \\
    \midrule
    \textbf{\method}     & \best{0.0620} & \best{0.0332} & ---     & \best{0.0472} & \best{0.0250} & ---     & \best{0.0953} & \best{0.0512} & ---     \\
    \midrule
    \multicolumn{10}{l}{\emph{G1: Head Architecture}} \\
    \quad w/o prefix conditioning        & 0.0175 & 0.0092 & $-72.0\%$ & 0.0160 & 0.0086 & $-65.5\%$ & 0.0352 & 0.0182 & $-64.3\%$ \\
    \quad summed prefix embeddings       & 0.0603 & 0.0324 & $-2.4\%$  & 0.0457 & 0.0243 & $-2.8\%$  & 0.0950 & 0.0503 & $-1.8\%$ \\
    \quad shared MLP head                & 0.0564 & 0.0301 & $-9.3\%$  & 0.0423 & 0.0225 & $-10.0\%$ & 0.0908 & 0.0481 & $-6.1\%$ \\
    \quad cascaded hidden state          & 0.0613 & 0.0328 & $-1.2\%$  & 0.0457 & 0.0245 & $-2.0\%$  & 0.0945 & 0.0503 & $-1.8\%$ \\
    \midrule
    \multicolumn{10}{l}{\emph{G2: Context Module}} \\
    \quad w/o multi-head attention            & 0.0597 & 0.0316 & $-4.8\%$  & 0.0442 & 0.0234 & $-6.4\%$  & 0.0929 & 0.0496 & $-3.1\%$ \\
    \quad per-digit context readout      & 0.0618 & 0.0330 & $-0.4\%$  & 0.0467 & 0.0249 & $-0.3\%$  & 0.0959 & 0.0515 & $+0.7\%$ \\
    \quad w/o context FFN                & 0.0616 & 0.0330 & $-0.7\%$  & 0.0468 & 0.0248 & $-0.8\%$  & 0.0953 & 0.0507 & $-1.0\%$ \\
    \midrule
    \multicolumn{10}{l}{\emph{G3: Distillation Design}} \\
    \quad w/o KL                         & 0.0602 & 0.0323 & $-2.7\%$  & 0.0449 & 0.0240 & $-4.0\%$  & 0.0900 & 0.0478 & $-6.6\%$ \\
    \quad full-vocab logits              & 0.0619 & 0.0330 & $-0.6\%$  & 0.0470 & 0.0250 & $-0.0\%$   & 0.0956 & 0.0510 & $-0.5\%$ \\
    \quad w/o teacher embeddings         & 0.0616 & 0.0330 & $-0.6\%$  & 0.0467 & 0.0248 & $-0.8\%$  & 0.0955 & 0.0511 & $-0.2\%$ \\
    \bottomrule
  \end{tabular}}
  \renewcommand{\arraystretch}{1.0}
  \vspace{-1.0mm}
\end{table}

\textbf{G1: Head architecture.}
The variant without prefix conditioning predicts all four digits from the same context vector in parallel, without feeding prefix embeddings into later heads. It loses 64--72\% NDCG@10, showing that the prediction must condition on the ordered SID prefix. Summing prefix embeddings instead of concatenating them loses 1.8--2.8\%, so the heads benefit from seeing the prefix as an ordered tuple. On the summed-prefix input, sharing one classifier across all digits drops by 6.1--10.0\%, suggesting that different digits require dedicated heads. The cascaded-state variant passes the newest prefix token and the hidden activation to the next head, instead of giving every head the fixed context and the full prefix. It does not help, so direct access to the cached context and explicit prefix is better for four-digit SIDs.

\textbf{G2: Context module.}
The variant without multi-head attention replaces the one-shot attention readout with a linear projection of the mean-pooled encoder states. It loses 3.1--6.4\% NDCG@10, confirming that \method still needs token-level history alignment. A per-digit prefix-conditioned context readout changes NDCG@10 by $-0.4\%$ to $+0.7\%$, so repeated cross-attention at each step provides no consistent gain, confirming our observation that attention becomes redundant in SID-based GR. Removing the FFN after multi-head attention costs at most 1.0\%, indicating that the attention readout carries most of the useful context.

\textbf{G3: Distillation design.}
Removing KL distillation and training only on hard SID labels loses 2.7--6.6\%, showing that the teacher's dense soft labels help in the sparse valid-SID space. Predicting the full 1027-token vocabulary gives no benefit over the 256-way digit codebook. Replacing the frozen teacher embeddings with randomly initialized trainable embeddings changes NDCG@10 by at most 0.8\%, so the teacher embeddings are useful but not the main source of \method's accuracy.

\vspace{-0.4\baselineskip}
\subsection{Performance of \method under Different Settings}
\label{sec:exp-further}
\label{sec:exp-robustness}
\vspace{-0.2\baselineskip}

We evaluate \method under three settings: SID tokenizer, batch size, and beam size (Figure~\ref{fig:robustness}). A temporal item-shift diagnostic is reported in Appendix~\ref{app:temporal-shift}.

\textbf{Tokenizer sensitivity.}
We distill \method from teachers trained with popular tokenizers: RQ-KMeans~\citep{onerec,grid}, RQ-VAE~\citep{rqvae,tiger}, and PSID~\citep{psid}.  As shown in Figure~\ref{fig:robustness}(a), \method retains 99.6\% to 102.9\% of teacher NDCG@10 across all tokenizer--dataset pairs, with no systematic loss. 

\textbf{Batch size.}
Figure~\ref{fig:robustness}(b) plots throughput and peak GPU memory as batch size scales from 8 to 512.
TIGER-kv saturates early because each digit step still runs the Transformer decoder over all active beams.
\method instead turns the post-context computation into batched dense projections. It's throughput scales until batch size reaches $256$, where speedup peaks at $16.6\times$.
At batch size $512$, \method keeps peak memory below $1.5$\,GB, compared with up to $30$\,GB for TIGER-kv.

\textbf{Beam size.}
In Figure~\ref{fig:robustness}(c), TIGER’s throughput drops by 37\% as beam size grows from 10 to 50, due to repeated autoregressive decoding.
\method avoids this bottleneck: after the one-shot context readout, larger beams merely widen the batched MLP projections without adding sequential steps, maintaining stable throughput and outperforming the teacher's NDCG@10 across all beam sizes.

\begin{figure}[!t]
  \centering
  \begin{subfigure}[b]{0.32\linewidth}
    \centering
    \includegraphics[width=\linewidth]{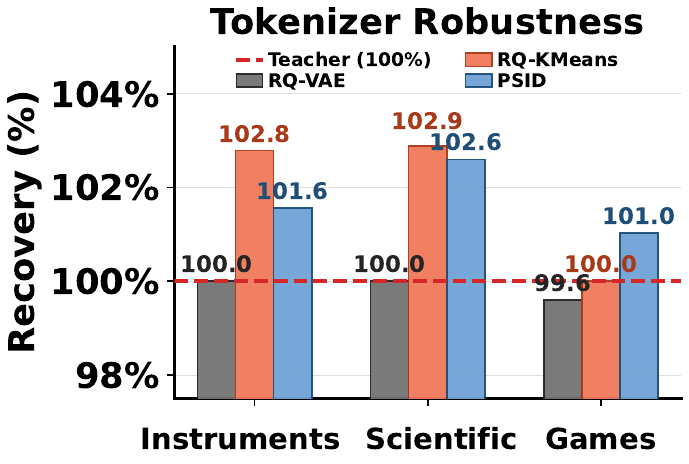}
    \caption{Tokenization strategies.}
    \label{fig:robustness-tok}
  \end{subfigure}\hfill
  \begin{subfigure}[b]{0.32\linewidth}
    \centering
    \includegraphics[width=\linewidth]{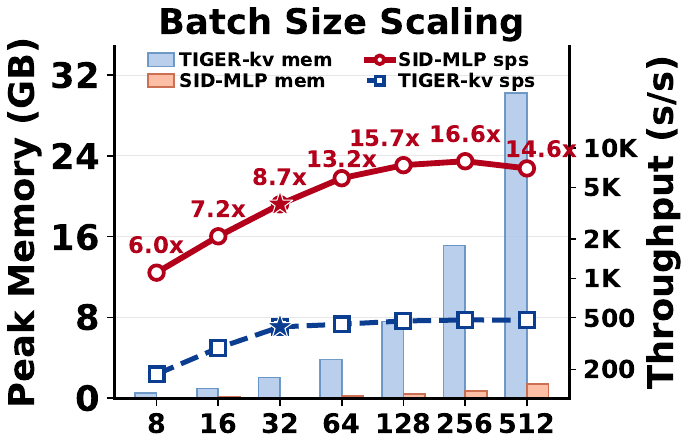}
    \caption{Batch size.}
    \label{fig:robustness-bs}
  \end{subfigure}\hfill
  \begin{subfigure}[b]{0.32\linewidth}
    \centering
    \includegraphics[width=\linewidth]{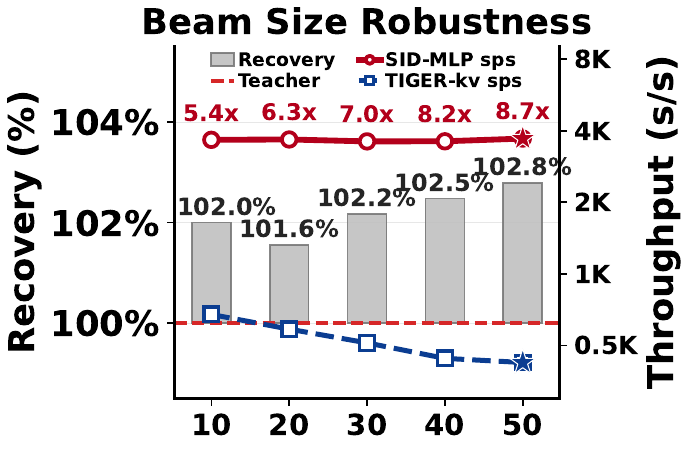}
    \caption{Beam size.}
    \label{fig:robustness-beam}
  \end{subfigure}
  \caption{\textbf{\method robustness across settings.} (a) NDCG@10 recovery is the ratio between \method and its teacher across tokenizers. (b) Peak GPU memory (bars) and throughput (lines) as batch size changes. (c) NDCG@10 recovery (bars) and throughput (lines) as beam size changes.}
  \vspace{-8pt}
  \label{fig:robustness}
\end{figure}

\vspace{-6pt}

\section{Related Work}

\label{sec:related}
\vspace{-4pt}
\textbf{Generative recommendation with identifiers.}
Generative retrieval stores items as discrete identifiers and retrieves by generating those identifiers autoregressively, as in DSI~\citep{dsi}. TIGER~\citep{tiger} brings this idea to recommendation with RQ-VAE semantic IDs and a T5 encoder--decoder; prompt, LLM, and large-scale GR systems extend backbones or serving regimes~\citep{p5,gptrec,lcrec,hstu,onerec,plum,liger}. SID studies mainly revise identifiers: indexing and semantic/collaborative tokenization~\citep{p5cid,hou2024bridging,eager,cost,simcit,cofirec,ccfrec,psid}, context-aware or multimodal IDs~\citep{actionpiece,pctx,mbgen,mmgrec,mql4grec,macrec,beyond_multimodal}, end-to-end or adaptive SID learning~\citep{letter,etegrec,bloger,pit,diger,unigrec,unisid, chen2024enhancing}, and analysis or deployment recipes~\citep{spmsid,grid,gr_scaling,gr_generalization,latte,snap_sid,sidstaleness,lee2026sequential}. These methods mostly change how IDs are built, analyzed, or how the recommender is trained. \method is complementary: given a trained TIGER-style tokenizer and encoder, it targets the repeated Transformer-decoder computation inside constrained beam search.

\textbf{Efficient generation for recommender IDs.}
Existing GR accelerators reduce latency in different ways. AtSpeed~\citep{atspeed} and SpecGR~\citep{specgr} draft candidate IDs and still use the target generative recommender for verification. NEZHA~\citep{nezha} adds self-drafting heads and a hash-based validity check, while EARN~\citep{earn} compresses LLMRec context with register tokens. Other efficient or alternative-generation designs use collaborative tokenization, compact modeling, parallel decoding, non-autoregressive reranking, or diffusion/masked-generation pipelines~\citep{setrec,rpg,nar4rec,marius,diffgrm,contrec,maskgr,lladarec}. These approaches improve serving through verification, architecture changes, identifier redesign, or modified generation procedures. \method instead keeps the existing 4-digit SID space and replaces online decoder calls with prefix-conditioned MLP heads, without target-model verification.

\textbf{Distillation and lightweight sequence models.}
Knowledge distillation transfers teacher behavior to smaller students~\citep{hinton,fitnets}; in recommendation and retrieval it is usually used for quality transfer to discriminative recommenders or generative retrievers~\citep{dllm2rec,dgr}. GLNN~\citep{glnn} and LLP~\citep{llp} show that an MLP student can absorb a teacher's structural inductive bias while removing inference-time dependencies. LLM inference accelerators span speculative decoding, KV-cache serving, and draft heads~\citep{specdec,shazeer2019fast,pagedattention,medusa,hydra,eagle,eagle3}; they target general text decoding or still require verification. State-space sequence models~\citep{s4,mamba,mamba2,gdn} provide efficient recurrent backbones, but do not directly exploit the low-branching SID prefix structure as prefix-conditioned MLP heads do.

\vspace{-4pt}

\section{Conclusion and Future Works}
\label{sec:conclusion}

\vspace{-7pt}

In this paper, we propose \method to eliminate the inference latency of autoregressive generative recommendation. Instead of relying on a heavy Transformer decoder, our key idea is to distill the generative process into cascaded, lightweight MLP heads. By computing the multi-head attention context exactly once, \method reduces beam expansion to efficient batched dense MLP operations without requiring repeated state/cache update. Extensive experiments demonstrate that \method matches the teacher's performance, achieving an $8.74\times$ throughput speedup and a 95.7\% peak-memory reduction. We also introduce \methodpp, an encoder-distilled variant that achieves a $10.25\times$ speedup. In future work, we plan to apply \method to industrial-scale systems with millions of items to further investigate search-space dynamics and encoder-side efficiency.

\bibliographystyle{unsrtnat}
\bibliography{references}

\appendix

\newpage
\section{Notations}
\label{app:notation}

\Cref{tab:notation} summarizes the notations used throughout the paper.

\begin{table}[htbp]
  \centering
  \small
  \caption{Notations and explanations.}
  \label{tab:notation}
  \setlength{\tabcolsep}{6pt}
  \renewcommand{\arraystretch}{1.0}
  \begin{tabular}{@{}p{0.20\textwidth}p{0.78\textwidth}@{}}
    \toprule
    \textbf{Notation} & \textbf{Explanation} \\
    \midrule
    $\mathcal{U}, \mathcal{V}$ & User and item sets; $u$ and $v$ index users and items \\
    $\mathbf{X}_u=[v_1,\ldots,v_n]$ & Historical item sequence of user $u$ \\
    $\mathbf{s}_u, S_u$ & Serialized encoder tokens for $\mathbf{X}_u$ and its non-padding length \\
    $c^{(v)}=(c_1,\ldots,c_L)$ & Semantic ID of item $v$; $c_t$ is digit $t$ and $c_{<t}$ its prefix \\
    $\mathcal{C}, C, L$ & Per-digit codebook, codebook size, and SID length; TIGER uses $C=256$ and $L=4$ \\
    $p_\theta(\cdot)$ & Teacher autoregressive distribution over SID digits \\
    $B, K, m$ & Beam size, recommendation cutoff, and teacher-generated prefix length in $m$-mode \\
    \midrule
    $E_T(\cdot), G_\phi(\cdot)$ & Frozen TIGER encoder and \methodpp student encoder; both map serialized user-history tokens to encoder states \\
    $\mathbf{H}_u, \hat{\mathbf{H}}_u$ & Teacher states and \methodpp encoder states \\
    $d_h, d_e$ & Encoder hidden-state size and frozen token-embedding size \\
    $e(\cdot)$ & Frozen SID-token embedding lookup \\
    $W_q, \mathbf{q}, \tilde{\mathbf{z}}, \mathbf{z}$ & Query projection/query and cached one-shot context vectors \\
    $f_t(\cdot)$ & Digit-$t$ MLP head mapping the prefix-conditioned input to sliced student logits \\
    $\mathbf{p}_t, \boldsymbol{\ell}_t$ & Input vector and student logits for digit $t$ \\
    $\tilde{\boldsymbol{\ell}}_t, c_t^\star$ & Teacher logits and ground-truth codebook index for digit $t$ \\
    \midrule
    $\alpha, \tau, \mathcal{L}_t, \mathcal{L}$ & Distillation weight, temperature, per-digit loss, and total loss \\
    $\mathcal{B}_t$ & Active beam set after digit $t$ \\
    $N, i, r$ & Teacher decoder layer count, encoder token position, and \methodpp layer index \\
    \midrule
    $\mathbf{x}_{u,i}^{(r)}, \mathbf{g}_u^{(r)}$ & \methodpp token state and pooled global vector at encoder layer $r$ \\
    $a_i, F_{a_i}(\cdot)$ & Digit-role id for token position $i$ and the corresponding role-specific MLP \\
    \bottomrule
  \end{tabular}
  \renewcommand{\arraystretch}{1.0}
\end{table}

\section{Motivation Details}
\subsection{Per-Dataset Codebook Branching Factor Statistics}
\label{app:codebook}

\Cref{tab:codebook} reports the per-dataset branching factors behind the search-space collapse discussed in \Cref{sec:motivation-perdigit}. In each evaluated catalog, the first digit has the full 256-way choice, the second digit has about 38 valid continuations on average, the third digit has about 2.2, and the fourth digit is close to singleton. Later digits still have occasional multi-choice prefixes, but their average search space is already near singleton within each dataset.

\begin{table}[!t]
  \caption{\textbf{Average codebook branching factor by digit.} Number of valid continuations at digit $t$ given a valid prefix $c_{<t}$, computed separately within each dataset's item vocabulary. No cross-dataset average is used.}
  \label{tab:codebook}
  \centering
  \small
  \begin{tabular}{lccccc}
    \toprule
    Dataset & \# items & Avg. $|c_1|$ & Avg. $|c_2\!\mid\! c_1|$ & Avg. $|c_3\!\mid\! c_{1:2}|$ & Avg. $|c_4\!\mid\! c_{1:3}|$ \\
    \midrule
    Instruments & 24{,}587 & 256 & 37.7 & 2.19 & 1.17 \\
    Scientific & 25{,}848 & 256 & 37.7 & 2.24 & 1.19 \\
    Games & 25{,}612 & 256 & 38.1 & 2.19 & 1.20 \\
    \bottomrule
  \end{tabular}
\end{table}

\subsection{Distillation Versus Scratch Training}
\label{app:scratch}

We compare distillation against scratch-training to justify our focus on the distilled setting. \Cref{tab:scratch-distill} evaluates a 1-layer Transformer decoder ($m{=}0$) and our \method decoder on Instruments dataset. For scratch-training, we train exclusively on hard SID labels by removing the KL term.

\begin{table}[!t]
  \centering
  {\setlength{\tabcolsep}{5.5pt}\renewcommand{\arraystretch}{0.95}\small
  \caption{\textbf{Distillation versus scratch-training on Instruments.} Values are NDCG@10. The relative drop is measured against the distillation row.}
  \label{tab:scratch-distill}
  \begin{tabular}{lccc}
    \toprule
    Decoder Family & Distill & Scratch & Drop vs. Distill \\
    \midrule
    1-layer Transformer decoder & 0.0318 & 0.0313 & $-1.6\%$ \\
    MLP decoder (\method) & 0.0332 & 0.0323 & $-2.7\%$ \\
    \bottomrule
  \end{tabular}\par}
\end{table}

Without distillation, the 1-layer Transformer loses 1.6\% NDCG@10. However, for the simpler MLP decoder, this gap widens to 2.7\%. This confirms that as the architecture becomes lighter, it relies more heavily on the teacher's soft labels to maintain accuracy. Therefore, in this work, we focus on distillation to achieve MLP-level efficiency without severe performance degradation.
\subsection{Cross-Dataset Attention Ablation}
\label{app:nmode}

\Cref{tab:nmode-full} checks whether the attention-ablation pattern in \Cref{tab:attn} is specific to Instruments dataset. We repeat the $m{=}0$ 1-layer decoder ablation on all three datasets. Removing self-attention causes a modest loss, while removing cross-attention causes a large loss on every dataset. This supports the design choice in \method: retain an encoder-conditioned context readout and remove repeated decoder attention.

\begin{table}[!t]
  \centering
  {\setlength{\tabcolsep}{6pt}\renewcommand{\arraystretch}{0.95}\small
  \caption{\textbf{Cross-dataset attention ablation on the $m{=}0$ 1-layer decoder.} SA and CA denote self-attention and cross-attention, respectively. Values are N@10; relative drops are computed against the first row within each dataset.}
  \label{tab:nmode-full}
  \begin{tabular}{ccccc}
    \toprule
    SA & CA & Instruments & Scientific & Games \\
    \midrule
    \(\checkmark\) & \(\checkmark\) & 0.0318 & 0.0236 & 0.0480 \\
    \(\times\) & \(\checkmark\) & 0.0295 ($-7.2\%$) & 0.0205 ($-13.1\%$) & 0.0449 ($-6.5\%$) \\
    \(\checkmark\) & \(\times\) & 0.0130 ($-59.1\%$) & 0.0088 ($-62.7\%$) & 0.0123 ($-74.4\%$) \\
    \(\times\) & \(\times\) & 0.0111 ($-65.1\%$) & 0.0072 ($-69.5\%$) & 0.0095 ($-80.2\%$) \\
    \bottomrule
  \end{tabular}\par}
\end{table}

\section{Experimental Setup and Reproducibility}
\subsection{Dataset Statistics}
\label{app:datasets}

\Cref{tab:datasets} summarizes the Amazon Reviews 2023~\citep{hou2024bridging} datasets used in the main experiments. User sequences are chronologically ordered, truncated to 20 items, and split via the leave-last-out protocol. 

\begin{table}[!t]
  \caption{\textbf{Dataset statistics (Amazon Reviews 2023).} Computed from our 5-core preprocessed splits (McAuley-Lab/amazon-reviews-2023, \texttt{last\_out} split). Four training targets per user sequence (rolling next-item); evaluation uses leave-one-out.}
  \label{tab:datasets}
  \centering
  \small
  \begin{tabular}{lrrrrc}
    \toprule
    Dataset & \# Users & \# Items & \# Interactions & Sparsity & Avg.\ Seq.\ Len \\
    \midrule
    Musical Instruments      & 57{,}439 & 24{,}587 & 511{,}836 & 99.96\% & 8.91 \\
    Industrial \& Scientific & 50{,}985 & 25{,}848 & 412{,}947 & 99.97\% & 8.10 \\
    Video Games              & 94{,}762 & 25{,}612 & 814{,}586 & 99.97\% & 8.60 \\
    \bottomrule
  \end{tabular}
\end{table}

\subsection{Implementation Details and Hyperparameters}
\label{app:hyperparams}

This subsection details the training configurations deferred from~\Cref{sec:exp-setup}.

\textbf{\method training.}
We optimize the trainable modules using AdamW with a cosine learning-rate schedule. Offline student training takes approximately one hour on average for each dataset (Instruments, Scientific, and Games). This offline cost is separate from online serving and is excluded from throughput measurements. Throughput profiling uses the trained checkpoints under the protocol in Appendix~\ref{app:hardware-profile}.

\textbf{Hyperparameter sweep.}
We sweep learning rate $\in\{1\mathrm{e}{-}5, 5\mathrm{e}{-}5, 1.25\mathrm{e}{-}4, 2.5\mathrm{e}{-}4, 5\mathrm{e}{-}4\}$, dropout $\in\{0.1, 0.2\}$, weight decay $\in\{0, 1\mathrm{e}{-}4, 1\mathrm{e}{-}3\}$, distillation weight $\alpha \in\{0, 0.3, 0.5, 0.7, 0.8, 1.0\}$, attention heads $\in\{4, 8\}$, attention inner dimension $\in\{128, 256, 384, 512\}$, and head hidden size $\in\{128, 256, 512, 768, 1024, 1536, 2048\}$.

\textbf{\methodpp training.}
For the encoder distillation (\Cref{sec:method-enc}), we instantiate the position-specific MLP encoder with 4 layers. Both Stage 1 (MSE pre-training) and Stage 2 (logit distillation) utilize the same AdamW optimizer and hyperparameter search space as \method.

\textbf{Hyperparameter sweep.}
We use a targeted \methodpp sweep centered on the best per-dataset \method settings. For Stage 1, we sweep \methodpp depth \(\in\{1,2,4,6,8\}\), FFN dimension \(\in\{256, 512, 1024, 2048, 4096, 8192\}\) with dataset-local candidates (e.g., larger dimensions up to 8192 for Games), learning rate \(\in\{5\mathrm{e}{-}4, 1\mathrm{e}{-}3\}\). For Stage 2, we freeze the encoder and train the decoder side \method using the same hyperparameter scope as introduced before.

\subsection{Hardware Profiling}
\label{app:hardware-profile}

\textbf{Reproducibility setup.}
All experiments run on a single NVIDIA RTX A6000 GPU (48GB) using PyTorch 2.1, CUDA 11.8, and Python 3.10. 

\textbf{Profiling protocol.}
\Cref{tab:hardware-profile-mi} details the runtime metrics behind the main results in \Cref{tab:main}. The timed inference region starts after model initialization and data loading, encompassing GPU forward passes, beam expansion, valid-SID masking, final item lookup, and metric computation. Offline teacher feature extraction for \method training is excluded from this online serving profile. We use batch size 32 and beam size $B{=}50$ across all tested methods. Throughput and speedup are averaged across the three datasets, while NDCG@10 and peak memory are reported from the Musical Instruments run (memory footprints remain stable across datasets). 

\textbf{Metric definitions.}
\emph{NDCG@10} measures ranking quality. \emph{samples/s} denotes end-to-end throughput, and \emph{ms/sample} is its inverse ($1000/\mathrm{samples\mbox{/}s}$). \emph{Speedup} is relative to TIGER-kv. \emph{Peak Mem.} captures the maximum allocated GPU memory during inference, and \emph{Mem./TIGER-kv} reports this footprint as a percentage of the TIGER-kv baseline. 

\begin{table*}[t]
  \caption{\textbf{Hardware profile.} Throughput, speedup, and latency are averaged across the three datasets; NDCG@10 and peak GPU memory are measured on Instruments. TIGER-kv serves as the speedup and memory-ratio baseline. \best{Bold} = best and \second{underline} = second-best per metric.}
  \label{tab:hardware-profile-mi}
  \centering
  \small
  \setlength{\tabcolsep}{3.2pt}
  \renewcommand{\arraystretch}{0.95}
  \begin{tabular}{lrrrrrr}
    \toprule
    Method & NDCG@10 & samples/s & Speedup & ms/sample & Peak Mem. (GB) & Mem./TIGER-kv \\
    \midrule
    TIGER & 0.0323 & 320 & 0.75$\times$ & 3.13 & 0.84 & 40.6\% \\
    TIGER-kv & 0.0323 & 424 & 1.00$\times$ & 2.36 & 2.07 & 100.0\% \\
    NEZHA & 0.0308 & 3{,}082 & 7.27$\times$ & 0.32 & \best{0.07} & \best{3.4\%} \\
    EARN & 0.0319 & 822 & 1.94$\times$ & 1.22 & 0.21 & 10.1\% \\
    Mamba2 & 0.0326 & 517 & 1.22$\times$ & 1.93 & 2.55 & 123.2\% \\
    GDN & 0.0321 & 510 & 1.20$\times$ & 1.96 & 9.67 & 467.1\% \\
    AtSpeed-S & 0.0323 & 286.3 & 0.68$\times$ & 3.49 & 0.87 & 42.0\% \\
    AtSpeed-R & 0.0204 & 94.0 & 0.22$\times$ & 10.64 & 0.93 & 44.9\% \\
    \midrule
    \rowcolor[gray]{0.93}\textbf{\method} & \best{0.0332} & \second{3{,}706} & \second{8.74$\times$} & \second{0.27} & \second{0.09} & \second{4.3\%} \\
    \rowcolor[gray]{0.93}\textbf{\methodpp} & \second{0.0327} & \best{4{,}347} & \best{10.25$\times$} & \best{0.23} & 0.43 & 20.8\% \\
    \bottomrule
  \end{tabular}
  \renewcommand{\arraystretch}{1.0}
\end{table*}

The profile reveals a practical Pareto frontier. \method achieves the best NDCG@10 while remaining 8.74$\times$ faster than TIGER-kv end-to-end and reducing peak memory from 2.07GB to a mere 0.09GB (a 95.7\% reduction). When isolating the generation phase, \method yields a \textbf{16.25$\times$ decoder-only speedup} (13{,}130 vs. 808 samples/s for TIGER-kv). \methodpp pushes the end-to-end throughput further to 4{,}347 samples/s (10.25$\times$ speedup) with a minor accuracy trade-off. While NEZHA has a slightly smaller absolute memory footprint (0.07GB), \method is substantially faster and more accurate. In contrast, SSM baselines (Mamba2 and GDN) incur much larger memory footprints and lower throughput, confirming that maintaining linear recurrent states is a poor latency-memory trade-off for short SID beam search.
\subsection{Random Seeds and Teacher-Matching Significance}
\label{app:seeds}

This appendix provides the multi-seed stability analysis for \method. Our core claim is quality preservation: \method replaces the autoregressive decoder without degrading the accuracy of the original TIGER-kv checkpoint. \Cref{tab:seed-noninferiority} reports the seed-level NDCG@10 values and a formal non-inferiority test.

\textbf{Seed protocol.}
We train \method across three random seeds (\(\{42,43,44\}\)). The seeds control the model initialization, data loader order, and minibatch stochasticity. TIGER-kv is evaluated using the single fixed teacher checkpoint from the main paper, serving as a deterministic reference.

\textbf{Non-inferiority test.}
Since our objective is to match rather than outperform the teacher, we employ a one-sided non-inferiority \(t\)-test. Let \(x_i\) denote the \method NDCG@10 for seed \(i\), \(b\) denote the deterministic TIGER-kv baseline, and \(\epsilon = 0.01b\) define a 1\% relative non-inferiority margin. The hypotheses are:
\[
H_0: \mathbb{E}[x_i - b] \le -\epsilon,
\qquad
H_1: \mathbb{E}[x_i - b] > -\epsilon.
\]
The test statistic is \(t = (\bar{x} - b + \epsilon) / (s/\sqrt{n})\), with \(n=3\) and degrees of freedom \(\mathrm{df}=2\), where \(s\) is the sample standard deviation. A \(p\)-value \(< 0.05\) rejects the null hypothesis, formally confirming that \method does not incur a statistically significant performance drop beyond the 1\% margin. 

\begin{table}[t]
  \caption{\textbf{Random-seed stability and non-inferiority test on NDCG@10.} Values are computed over seeds \(\{42,43,44\}\). The symbol \(\dagger\) indicates passing the non-inferiority test against TIGER-kv with a 1\% relative margin (\(p<0.05\)).}
  \label{tab:seed-noninferiority}
  \centering
  \small
  \setlength{\tabcolsep}{4.4pt}
  \resizebox{\linewidth}{!}{%
  \begin{tabular}{lccc cccc}
    \toprule
    Dataset & N@10 (42) & N@10 (43) & N@10 (44) & Mean\(\pm\)std & TIGER-kv & Diff. & \(p\) \\
    \midrule
    Instruments & 0.0333 & 0.0332 & 0.0331 & 0.0332\(\pm\)0.0001\(\dagger\) & 0.0323 & +0.0009 & 0.001 \\
    Scientific  & 0.0252 & 0.0250 & 0.0248 & 0.0250\(\pm\)0.0002\(\dagger\) & 0.0243 & +0.0007 & 0.007 \\
    Games       & 0.0512 & 0.0512 & 0.0511 & 0.0512\(\pm\)0.0001\(\dagger\) & 0.0512 & +0.0000 & 0.006 \\
    \bottomrule
  \end{tabular}}
\end{table}

As shown in \Cref{tab:seed-noninferiority}, all three datasets strongly satisfy the \(p<0.05\) criterion. This multi-seed evaluation statistically grounds our claim: \method successfully preserves the teacher's ranking quality while delivering the substantial speedups reported in \Cref{tab:main}.

\section{Cross-Backbone and Baseline Reproductions}
\subsection{Application to LC-Rec}
\label{app:lcrec}

This appendix checks whether \method also applies to an LLM-based generative recommender. LC-Rec~\citep{lcrec} releases public checkpoints on Amazon Reviews 2018~\citep{ni2019justifying}, so we follow its protocol on Musical Instruments, Arts Crafts and Sewing \textbf{(Arts)}, and Video Games. The dataset statistics are shown in \Cref{tab:datasets-2018}.

\begin{table}[!t]
  \caption{\textbf{Dataset statistics (Amazon Reviews 2018).} Users and items with fewer than five interactions are filtered.}
  \label{tab:datasets-2018}
  \centering
  \small
  \begin{tabular}{lrrrrc}
    \toprule
    Dataset & \# Users & \# Items & \# Interactions & Sparsity & Avg.\ Seq.\ Len \\
    \midrule
    Musical Instruments   & 24{,}773 & 9{,}923  & 206{,}153 & 99.92\% & 8.32 \\
    Arts Crafts and Sewing                  & 45{,}142 & 20{,}957 & 390{,}832 & 99.96\% & 8.66 \\
    Video Games           & 50{,}547 & 16{,}860 & 452{,}989 & 99.95\% & 8.96 \\
    \bottomrule
  \end{tabular}
\end{table}

\begin{figure}[!t]
  \centering
  \includegraphics[width=\linewidth,trim=8 0 8 0,clip]{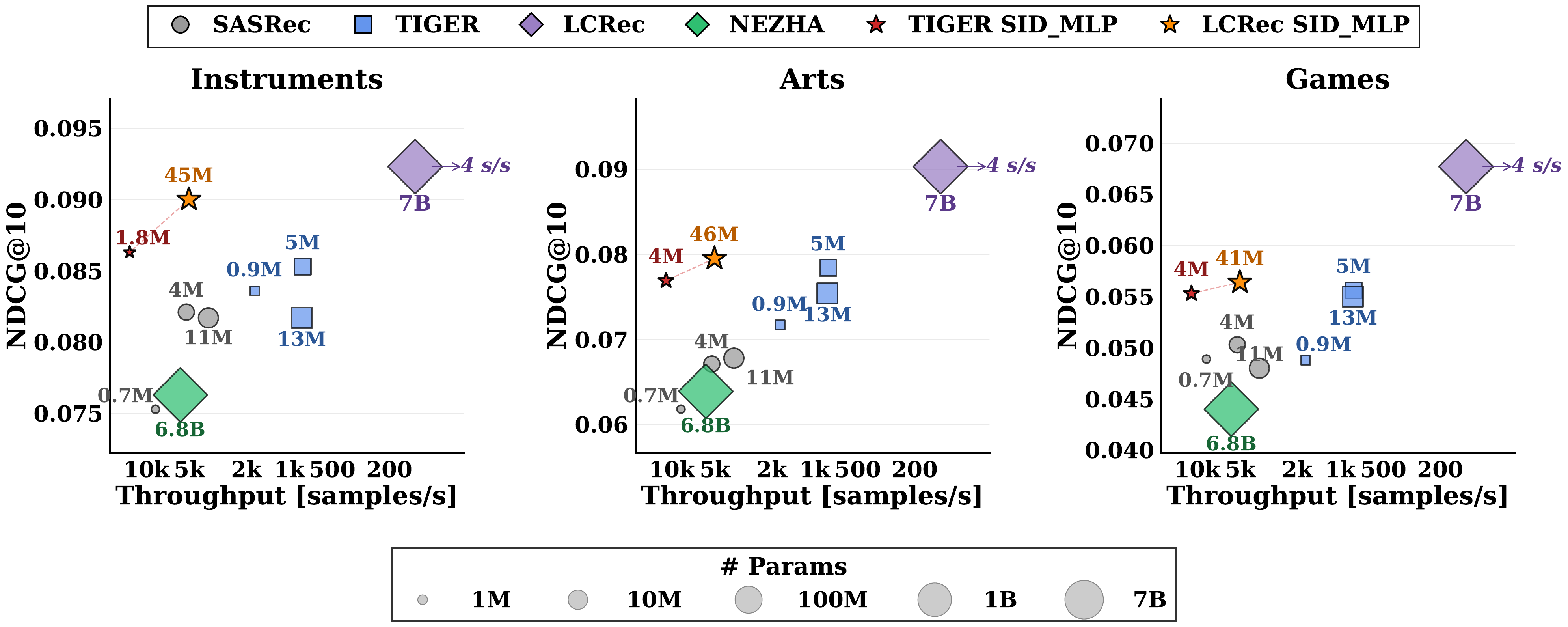}
  \caption{\textbf{Cross-scale Pareto on Amazon Reviews 2018.} NDCG@10 vs throughput (samples/s, log scale) across Instruments, Arts, and Games. All throughputs are end-to-end except LC-Rec 7B and LC-Rec \method, reported \emph{decode-only} (prefill excluded) as optimistic upper bounds. Baselines span SASRec~\citep{sasrec} (${\sim}$1M/5M/13M), TIGER (${\sim}$1M/5M/13M), and LC-Rec 7B (LLaMA-2, off-axis at ${\sim}4$\,samples/s); students are TIGER \method and LC-Rec \method, distilled from the 5M and 7B teachers respectively. Marker size encodes parameter count.}
  \label{fig:pareto}
\end{figure}

\textbf{Setup.}
Two design philosophies exist for accelerating LLM-based generative recommenders: (i) jointly fine-tune the LLM backbone with a draft head (e.g.\ NEZHA~\citep{nezha}), which breaks plug-and-play deployment; or (ii) keep the LLM frozen and train a plug-and-play student, preserving compatibility with existing checkpoints. \method follows (ii). We distill \method from \textbf{LC-Rec}, a 7B LLaMA-2-based generative recommender. Since LC-Rec only releases public checkpoints on Amazon Reviews 2018, we adopt its protocol; retraining TIGER at ${\sim}$1M/5M/13M on the same categories gives a single cross-scale Pareto view (\Cref{fig:pareto}).

\textbf{Adaptation to decoder-only LLaMA.}
LC-Rec has no separate encoder, requiring us to redesign the multi-head attention sources in \Cref{eq:global-context}. Recent studies on LLMRec identify a layer-wise attention sparsity inversion: early layers retain dense, informative patterns that encode broad user history, while later layers become highly sparse and redundant~\citep{earn,gems}. Building on this insight, we extract the key--value (KV) states from all history-item tokens at an early layer of the frozen prefill (e.g., the 8th of 32). For the query, we extract the last-layer (32nd) hidden state of the final prompt token, as it is directly optimized for next-token prediction and closely aligns with the immediate recommendation intent. The per-digit MLP heads, trie-constrained beam search, and KL+CE objective remain identical to the T5-based \method.

\textbf{\method training on LC-Rec.}
The training protocol follows the TIGER \method setup in Appendix~\ref{app:hyperparams}. To accommodate the LLaMA backbone, we simply project the 4096-dimensional Q and KV states down to $d_h{=}768$ via learnable linear layers. For hyperparameters, we sweep the number of stacked attention readout layers over \(\{1, 2, 3\}\), FFN dimension $\in\{512, 1024, 2048, 3072\}$, and head hidden size $\in\{512, 1024, 2048, 3072\}$.

\textbf{Findings.}
\method sits on the Pareto frontier at both teacher scales. At the TIGER scale, it matches the teacher while delivering an $8.74\times$ throughput speedup. At the 7B LC-Rec scale, \method recovers 83.3--97.5\% of the teacher's accuracy using only 44.1M trainable parameters ($0.6\%$ of the teacher). It accelerates end-to-end throughput to 38.04 samples/s ($17.9\times$ speedup) and decode-only throughput to 5{,}194 samples/s ($1{,}262.8\times$ speedup). The remaining accuracy gap to the frozen backbone varies by category. In contrast, NEZHA requires 6.8B trainable parameters ($153.2\times$ more than \method) yet only recovers 65.0--82.7\% of the LC-Rec teacher. As detailed in Appendix~\ref{app:nezha-repro}, our NEZHA reproductions consistently exhibit larger quality drops than \method across both architectures.

\subsection{AtSpeed Adaptation}
\label{app:atspeed-analysis}

\textbf{Adaptation target.}
AtSpeed is adapted to TIGER rather than used in its original open-vocabulary causal-LM setting. TIGER uses a T5 encoder--decoder and emits four constrained SID digits.

\textbf{Evaluation setting.}
The AtSpeed rows in \Cref{tab:main,tab:hardware-profile-mi} use the same TIGER checkpoints and sentence-T5 SIDs as the main results. The profile uses the full test split and beam \(B=50\). AtSpeed-S uses strict tree verification: an accepted step requires the teacher top-\(B\) beams to be contained in the draft tree. AtSpeed-R uses relaxed verification: draft samples are accepted by rejection sampling against the teacher distribution, and missing beams are filled from the correction distribution.

\textbf{Why AtSpeed is slow on TIGER.}
Original AtSpeed targets long autoregressive LM decoding. Its speedup comes from using a cheap draft to verify many future tokens of a costly teacher. TIGER does not match this setting. The teacher has only 4.59M parameters, and generation stops after four constrained SID digits. AtSpeed-S reaches 286.3 samples/s, or \(0.68\times\) TIGER-kv. It matches teacher quality because it verifies teacher greedy top-\(B\) containment, but it still pays for draft beam search and teacher tree verification. With only four decoder steps, this overhead is hard to amortize. AtSpeed-R reaches 94.0 samples/s, or \(0.22\times\) TIGER-kv and trails TIGER by 27--37\% NDCG@10 across the three datasets. This drop follows from the relaxed sampling target.

\subsection{EARN and State-Space Decoder Adaptations}
\label{app:baseline-adaptations}

This subsection documents the remaining baseline ports used in \Cref{tab:main}. These adaptations keep the TIGER SID tokenizer and valid-SID constraints fixed, so the comparison isolates decoder-side acceleration under the same recommendation task. Unless noted otherwise, baseline hyperparameters follow the authors' recommendations unless adaptation to the TIGER backbone is required.

\textbf{EARN adaptation.}
We adapt EARN to the TIGER encoder because TIGER stores the user history in encoder states before SID decoding. One prefix register and one suffix register are placed at the first and last valid encoder positions. After encoder layer \(k\), all non-register states are pruned; the remaining encoder layers and the decoder cross-attention use only the register states. We jointly fine-tune the resulting T5 model and the register embeddings, sweep \(k\in\{1,2,3\}\), and select checkpoints by validation NDCG@10.

\textbf{EARN attention diagnostic.}
We also check whether TIGER supports EARN's head/tail register placement. On 1,280 validation samples, TIGER encoder attention does not form the dual sink reported by EARN: across the four encoder layers, the first three positions receive 10.1--14.2\% of attention and the last three positions receive 11.0--11.7\%, close to or below the uniform three-position baseline of 13.6\%. The top-attended positions are mostly middle history tokens. This differs from the decoder-only premise used by EARN, where BOS is repeatedly visible through causal attention and the final prompt token can summarize the full history.

\textbf{State-space decoder adaptation.}
GDN and Mamba2 use the same frozen TIGER encoder features as \method. Each model receives the encoder hidden states followed by a learned start token or a bridge token and prefix embeddings from the frozen TIGER token embedding. It outputs 256-way logits for each SID digit and applies the same valid-SID constraints during beam search. We train only the SSM-side modules, bridge/start parameters, and output head with the same KL+CE objective as \method.
During beam search, each active prefix carries an SSM recurrent state; every digit step gathers parent states, updates them with the selected prefix token, and forks states for the expanded beams. This makes the short-SID setting dominated by state movement rather than long-sequence scan complexity.

\subsection{NEZHA Reproduction and Adaptation Details}
\label{app:nezha-repro}

This appendix documents how we reproduced NEZHA~\citep{nezha} in our two teacher settings. The key point is that we preserve NEZHA's hidden-state contract: a backbone pass produces one root hidden state and one hidden state per SID digit, and an MTP head predicts SID digits with per-position heads plus a recurrent state update and joint fine-tuning. The adaptation only maps NEZHA's placeholder construction to the teacher architecture: encoder-side placeholders for TIGER, response-side placeholders after the LC-Rec prompt for LC-Rec.

\textbf{TIGER adaptation.}
TIGER encodes the user sequence with a T5 encoder and normally decodes the target SID with a T5 decoder. To instantiate NEZHA without changing the SID tokenizer, we use the same abstract input pattern on the encoder side:
\[
  [\mathrm{BOS}, q_{\mathrm{user}}, x_1,\ldots,x_t,
  \mathrm{SP}_1,\mathrm{SP}_2,\mathrm{SP}_3,\mathrm{SP}_4],
\]
where \(q_{\mathrm{user}}\) is TIGER's user/context token and \(x_1,\ldots,x_t\) are the history SID tokens. The final context hidden state before \(\mathrm{SP}_1\) is used as $h_0$, and the four placeholder hidden states are used as $h_1,\ldots,h_4$. The T5 decoder is bypassed in the NEZHA path. The T5 encoder and the MTP head are jointly fine-tuned, matching NEZHA's joint backbone-plus-head training principle. The target remains exactly the four TIGER SID digits.
For TIGER NEZHA, we tuned the base learning rate over \(\{5\mathrm{e}{-}5, 1.25\mathrm{e}{-}4, 2.5\mathrm{e}{-}4, 5\mathrm{e}{-}4, 1\mathrm{e}{-}3\}\), swept weight decay over \(\{1\mathrm{e}{-}5, 1\mathrm{e}{-}3, 5\mathrm{e}{-}2\}\), and tested NEZHA's official parameter-group learning-rate scaling; neither improved the reported TIGER result.

\textbf{LC-Rec adaptation.}
LC-Rec is already a decoder-only LLaMA-based generative recommender trained with its own prompt template. For adaptation, the input is the same instruction prompt used by LC-Rec with an empty response field, followed by four SID placeholders:
\[
  [\mathrm{BOS}, q_{\mathrm{pre}}, x_1,\ldots,x_t, q_{\mathrm{resp}},
  \mathrm{SP}_1,\mathrm{SP}_2,\mathrm{SP}_3,\mathrm{SP}_4],
\]
where \(q_{\mathrm{pre}}\) and \(q_{\mathrm{resp}}\) are the LC-Rec prompt text before and after the rendered history; \(q_{\mathrm{resp}}\) includes the response header. Thus the placeholders are appended after LC-Rec's response prompt, not directly after the final historical item SID. The hidden state of the final prompt token before \(\mathrm{SP}_1\) is $h_0$, and the four placeholder hidden states are $h_1,\ldots,h_4$. The labels at the placeholder positions are the original four LC-Rec SID tokens. This keeps the LC-Rec teacher's input semantics intact: the history and instruction remain in the prompt, while the response SID span is replaced by NEZHA placeholders.

\textbf{LC-Rec NEZHA training.}
We fine-tuned the 7B LC-Rec backbone across four A6000 GPUs. Given the computational cost, we focused our sweep on the base learning rate \(\{1\mathrm{e}{-}6, 5\mathrm{e}{-}6, 1\mathrm{e}{-}5, 5\mathrm{e}{-}5, 1\mathrm{e}{-}4\}\) and selected the best validation checkpoint. To ensure a rigorous reproduction, we extensively tested other high-impact factors: we compared raw-SID inputs versus prompt-preserving inputs, applied NEZHA's official parameter-group learning rate multipliers (logit head $1{\times}$, token embedding $100{\times}$, transition $10{\times}$), and swept the inference MTP search budget over \(\{10, 20, 50, 512\}\). Despite these efforts, none of the configurations closed the accuracy gap to the LC-Rec teacher.

\textbf{Interpretation.}
On both TIGER and LC-Rec, NEZHA acts as a valid high-speed joint-finetuning baseline but consistently suffers from accuracy degradation. For TIGER, the port keeps NEZHA's MTP mechanism but moves placeholders to the encoder and trains with hard SID labels. For LC-Rec, the prompt-preserving adaptation isolates the modification strictly to the response generation path. Despite mechanically well-defined ports and extensive hyperparameter sweeps, NEZHA fails to automatically recover the quality of the original teachers.

\section{Additional Analyses}
\subsection{Hyperparameter and $m$-Mode Analysis}
\label{app:hparam-figs}

\begin{figure*}[!t]
  \centering
  \begin{subfigure}[b]{0.32\linewidth}
    \centering
    \includegraphics[width=\linewidth]{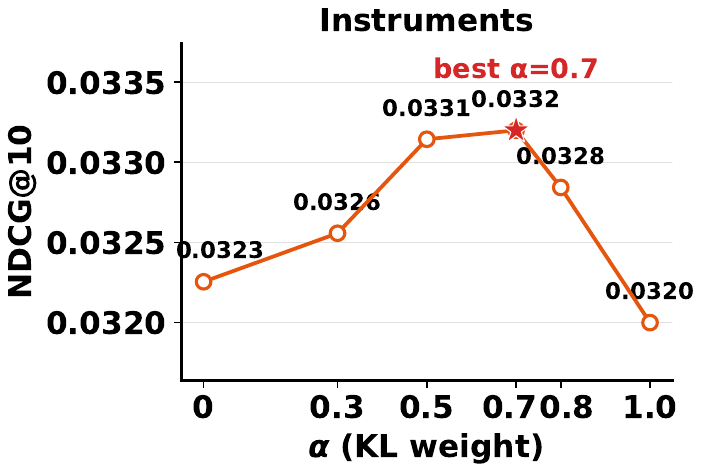}
    \caption{Instruments: $\alpha$.}
  \end{subfigure}\hfill
  \begin{subfigure}[b]{0.32\linewidth}
    \centering
    \includegraphics[width=\linewidth]{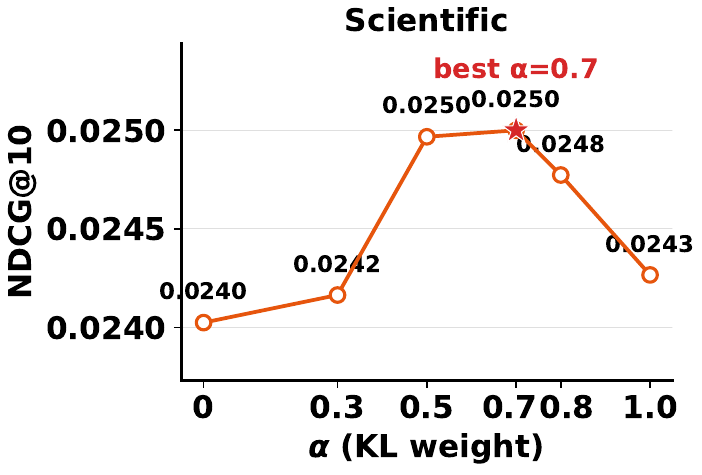}
    \caption{Scientific: $\alpha$.}
  \end{subfigure}\hfill
  \begin{subfigure}[b]{0.32\linewidth}
    \centering
    \includegraphics[width=\linewidth]{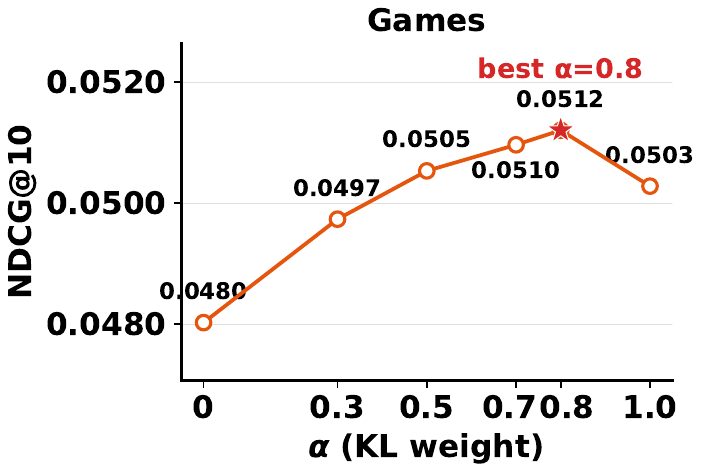}
    \caption{Games: $\alpha$.}
  \end{subfigure}

  \vspace{0.6em}

  \begin{subfigure}[b]{0.32\linewidth}
    \centering
    \includegraphics[width=\linewidth]{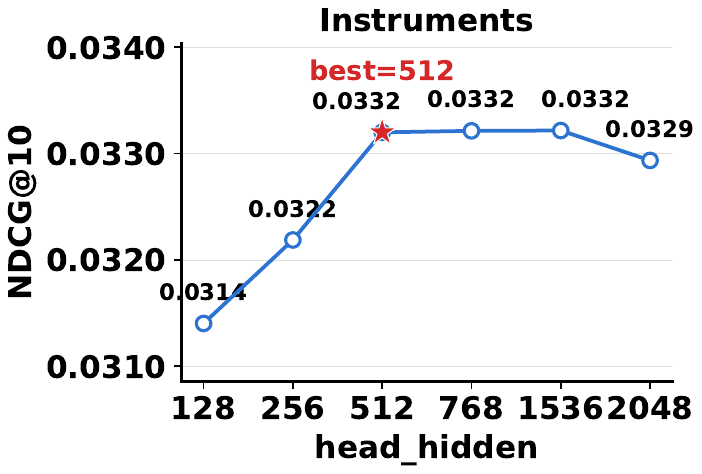}
    \caption{Instruments: head width.}
  \end{subfigure}\hfill
  \begin{subfigure}[b]{0.32\linewidth}
    \centering
    \includegraphics[width=\linewidth]{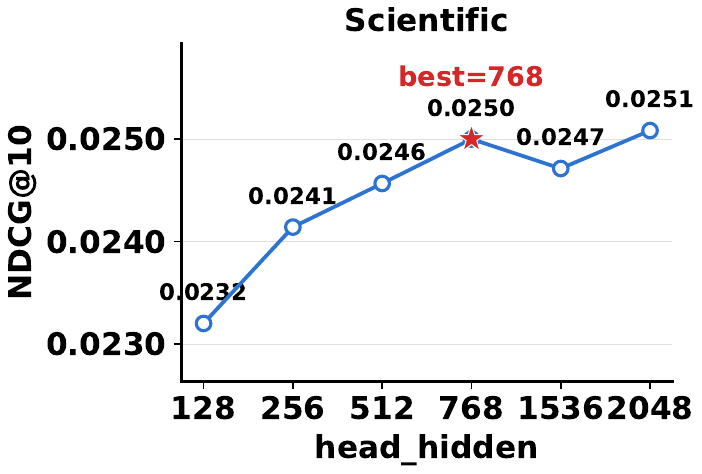}
    \caption{Scientific: head width.}
  \end{subfigure}\hfill
  \begin{subfigure}[b]{0.32\linewidth}
    \centering
    \includegraphics[width=\linewidth]{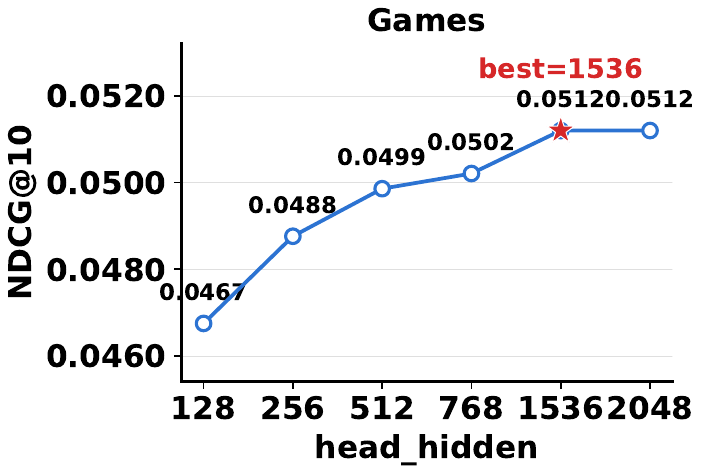}
    \caption{Games: head width.}
  \end{subfigure}

  \vspace{0.6em}

  \begin{subfigure}[b]{0.32\linewidth}
    \centering
    \includegraphics[width=\linewidth]{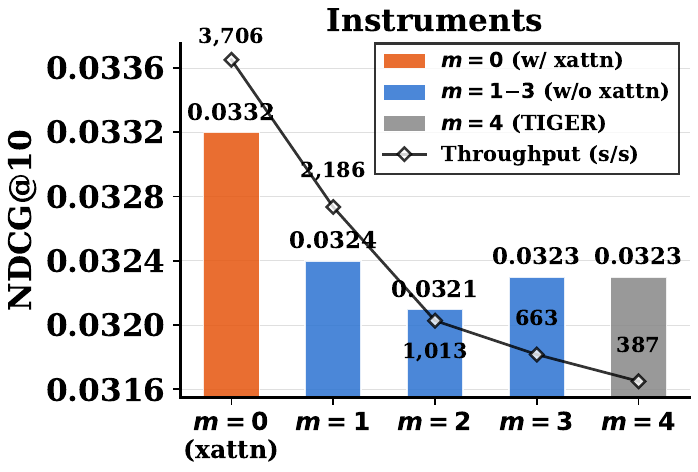}
    \caption{Instruments: $m$-mode.}
  \end{subfigure}\hfill
  \begin{subfigure}[b]{0.32\linewidth}
    \centering
    \includegraphics[width=\linewidth]{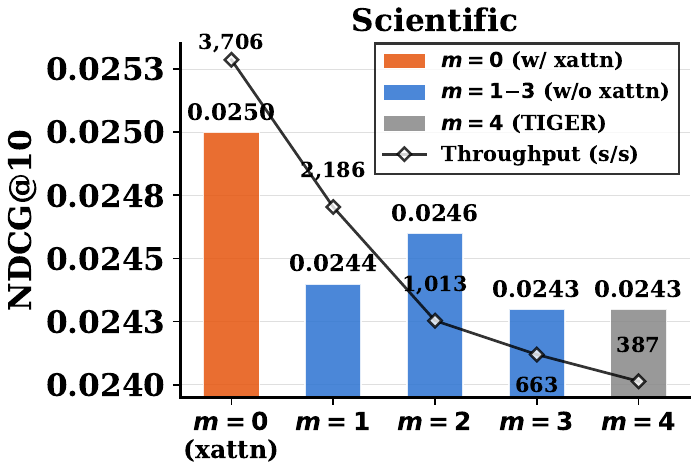}
    \caption{Scientific: $m$-mode.}
  \end{subfigure}\hfill
  \begin{subfigure}[b]{0.32\linewidth}
    \centering
    \includegraphics[width=\linewidth]{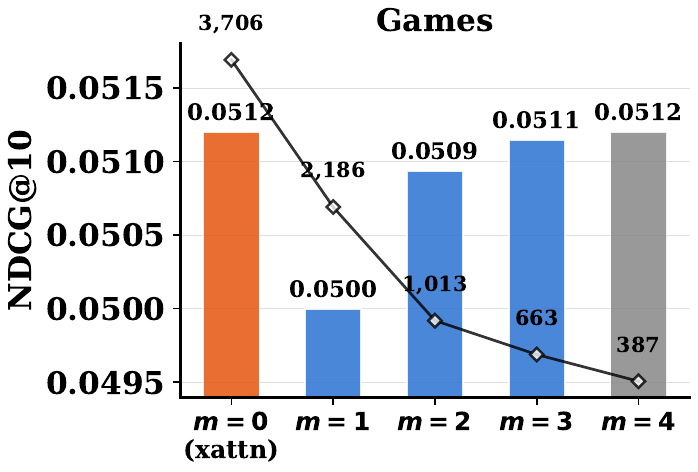}
    \caption{Games: $m$-mode.}
  \end{subfigure}

  \caption{\textbf{Hyperparameter and $m$-mode analysis.} Top row: $\alpha \in \{0,0.3,0.5,0.7,0.8,1.0\}$ sweep. Middle row: head-hidden width sweep. Bottom row: $m$-mode accuracy--throughput tradeoff. Columns correspond to Instruments, Scientific, and Games.}
  \label{fig:hparam-appendix}
\end{figure*}

This appendix reports the full hyperparameter and $m$-mode curves for Instruments, Scientific, and Games. \Cref{fig:hparam-appendix} is organized by analysis type: each row is one sweep, and each column is one dataset. All runs use the same distillation recipe as the main experiments.

\textbf{$\alpha$ sweep.}
Instruments and Scientific peak at $\alpha{=}0.7$, while Games peaks at $\alpha{=}0.8$. All three datasets show smooth curves with no sharp cliff, indicating that the KL+CE mixture is robust.

\textbf{Scale-up.}
NDCG@10 saturates by head hidden width 512 on Instruments, 768 on Scientific, and 1536 on Games. Throughput decreases monotonically with head width, so the saturation point gives the practical operating point for each dataset.

\textbf{$m$-mode tradeoff.}
All three datasets show the same tradeoff shape. $m{=}0$ is the fastest fully distilled setting and matches the teacher in the main results. $m{=}1$ anchors only the first SID digit with the teacher and provides an intermediate accuracy--speed point, consistent with the branching analysis in \Cref{sec:motivation-perdigit}.

\subsection{Distilled Encoder Ablations}
\label{app:encoder_ablation}

We ablate the \methodpp encoder distillation along two key design choices. \Cref{tab:encoder-distill-ablation} reports NDCG@10 across Instruments, Scientific, and Games; the $\Delta$ columns show the relative change versus the full \methodpp.

\textbf{Position-specific MLPs.} Sharing a single MLP across all positions consistently degrades NDCG@10 by 1.0--1.4\%. This indicates that using dedicated MLPs for different token roles is beneficial.

\textbf{Two-stage distillation.} The \emph{End2End} variant skips the Stage-1 hidden-state matching (MSE loss) and trains the encoder directly using only the final KL+CE ranking objective. This causes severe degradation: $-6.0\%$ on Instruments, $-11.3\%$ on Scientific, and $-6.6\%$ on Games. These substantial drops confirm that the final ranking loss is too sparse and indirect to optimize an MLP-based sequence encoder from scratch. Stage-1 MSE pre-training is crucial; it provides a dense, token-level supervision signal that aligns the lightweight encoder's representations before fine-tuning the prediction heads.

\begin{table}[!t]
  \caption{\textbf{\methodpp encoder distillation ablations.} Values are test NDCG@10 on Amazon Reviews 2023. $\Delta$ represents the relative change compared to the full \methodpp.}
  \label{tab:encoder-distill-ablation}
  \centering
  \small
  \setlength{\tabcolsep}{4.5pt}
  \begin{tabular}{lcccccc}
    \toprule
    & \multicolumn{2}{c}{\textbf{Instruments}} & \multicolumn{2}{c}{\textbf{Scientific}} & \multicolumn{2}{c}{\textbf{Games}} \\
    \cmidrule(lr){2-3}\cmidrule(lr){4-5}\cmidrule(lr){6-7}
    Variant & N@10 & $\Delta$ & N@10 & $\Delta$ & N@10 & $\Delta$ \\
    \midrule
    \textbf{\methodpp} & \textbf{0.0328} & --- & \textbf{0.0244} & --- & \textbf{0.0486} & --- \\
    \midrule
    Shared MLP & 0.0325 & $-1.0\%$ & 0.0242 & $-1.0\%$ & 0.0480 & $-1.4\%$ \\
    End2End & 0.0308 & $-6.0\%$ & 0.0217 & $-11.3\%$ & 0.0454 & $-6.6\%$ \\
    \bottomrule
  \end{tabular}
\end{table}
\subsection{Temporal Item-Shift Diagnostic}
\label{app:temporal-shift}

We evaluate \method's robustness to temporal item shift, where test targets represent recently introduced items. Following the timestamp-based evaluation protocols of SpecGR~\citep{specgr} and ColdGenRec~\citep{coldgenrec}, we compute each item's first interaction timestamp in the chronological log. A test case is assigned to the Temporal-80 or Temporal-90 subset if its target item's first interaction occurs after the global 80\% or 90\% timestamp, respectively. The 90\% cutoff strictly matches ColdGenRec's item cold-start definition, while the 80\% cutoff provides a broader diagnostic subset. Since this diagnostic uses the existing TIGER checkpoints, SID tokenizer, and candidate sets from the main experiments, it assesses temporal shift robustness rather than strict zero-shot item insertion.

\begin{table}[!t]
  \caption{\textbf{Temporal item-shift diagnostic.} Test NDCG@10 on targets whose first interaction timestamp is after the global 80\% or 90\% timestamp. $\Delta$ is \method minus TIGER.}
  \label{tab:temporal-shift}
  \centering
  \small
  \setlength{\tabcolsep}{3.4pt}
  \begin{tabular}{l rrrr rrrr}
    \toprule
    & \multicolumn{4}{c}{\textbf{Temporal-80}} & \multicolumn{4}{c}{\textbf{Temporal-90}} \\
    \cmidrule(lr){2-5}\cmidrule(lr){6-9}
    Dataset & \# Test & TIGER & \method & $\Delta$ & \# Test & TIGER & \method & $\Delta$ \\
    \midrule
    Instruments & 4{,}272  & 0.0076 & 0.0076 & +0.0000 & 2{,}051  & 0.0079 & 0.0078 & -0.0001 \\
    Scientific  & 5{,}573  & 0.0152 & 0.0158 & +0.0006 & 3{,}122  & 0.0134 & 0.0149 & +0.0015 \\
    Games       & 21{,}212 & 0.0193 & 0.0194 & +0.0001 & 10{,}434 & 0.0190 & 0.0191 & +0.0001 \\
    \midrule
    Average     & --       & 0.0140 & 0.0143 & +0.0003 & --       & 0.0134 & 0.0139 & +0.0005 \\
    \bottomrule
  \end{tabular}
\end{table}

\Cref{tab:temporal-shift} shows that \method seamlessly tracks the frozen TIGER teacher on temporally shifted targets. Across all three datasets, \method performs identically or slightly better, yielding average NDCG@10 improvements of $+0.0003$ and $+0.0005$ on the 80\% and 90\% splits, respectively. This confirms that replacing the autoregressive Transformer decoder with prefix-conditioned MLP heads introduces no measurable degradation on newer items.

\section{Broader Impact}
\label{app:impact}

Faster generative recommendation lowers compute and energy cost per recommendation, with positive environmental impact. No novel risks beyond generic recommender-system concerns (filter bubbles, engagement optimisation). It does not introduce a new recommendation objective, collect new user data, or change the user-modeling assumptions of the teacher recommender. Therefore, we do not introduce qualitatively new risks beyond those already associated with recommender systems.

\newpage

\end{document}